\documentclass[12pt,preprint]{aastex}

\newcommand{\etal  }{{et al.} }
\newcommand{\msun}{\thinspace M_\odot}  
\newcommand{\rsun}{\thinspace R_\odot}  
\newcommand{\vect}[1]{\mbox{\boldmath$#1$}}

\newcommand{\rhoc}{\rho_{\rm c}}
\newcommand{\nc}{n_{\rm c}}

\newcommand{\ob}{\varepsilon_{\rm ob}}
\newcommand{\ar}{\varepsilon_{\rm ar}}

\newcommand{\cm  }{\,{\rm cm}^{-3} } 

\newcommand{\dfrac}[2]{{\displaystyle \frac{#1}{#2}}  }
\newcommand{\km  }{\,{\rm km\, s^{-1}} } 
\newcommand{\ap}{ A_{ \varphi}}

\newcommand{\sep }{R_{\rm sep}}
\newcommand{\pb}{{\it b}}

\shorttitle{Wide and Close Binary Formation}
\shortauthors{Machida  \etal 2006}

\begin{document}
\title{Formation Scenario for Wide and Close Binary Systems}

\shortauthors{Machida  \etal 2007}

\author{Masahiro N. Machida\altaffilmark{1}, Kohji Tomisaka\altaffilmark{2}, Tomoaki Matsumoto\altaffilmark{3}, and  Shu-ichiro Inutsuka\altaffilmark{1}} 
\altaffiltext{1}{Department of Physics, Graduate School of Science, Kyoto University, Sakyo-ku, Kyoto 606-8502, Japan; machidam@scphys.kyoto-u.ac.jp, inutsuka@tap.scphys.kyoto-u.ac.jp}
\altaffiltext{2}{National Astronomical Observatory, Mitaka, Tokyo 181-8588, Japan, also School of Physical Sciences, Graduate University of Advanced Study (SOKENDAI); tomisaka@th.nao.ac.jp}
\altaffiltext{3}{Faculty of Humanity and Environment, Hosei University, Fujimi, Chiyoda-ku, Tokyo 102-8160, Japan; matsu@i.hosei.ac.jp}

\begin{abstract}
Fragmentation and binary formation processes are studied using three-dimensional resistive MHD nested grid simulations.
Starting with a Bonnor-Ebert isothermal cloud rotating in a uniform magnetic field, we calculate the cloud evolution from the molecular cloud core ($n=10^4\cm$) to the stellar core ($n\simeq 10^{22}\cm$), where $n$ denotes the central density.
We calculated 147 models with different initial magnetic, rotational, and thermal energies, and the amplitudes of the non-axisymmetric perturbation.
In a collapsing cloud, fragmentation is mainly controlled by the initial ratio of the rotational to the magnetic energy, regardless of the initial thermal energy and amplitude of the non-axisymmetric perturbation.
When the clouds have large rotational energies in relation to magnetic energies, fragmentation occurs in the low-density evolution phase ($10^{12}\cm \lesssim n \lesssim 10^{15}\cm$) with separations of 3--300\,AU.
Fragments that appeared in this phase are expected to evolve into wide binary systems.
On the other hand, fragmentation does not occur in the low-density evolution phase, when initial clouds have large magnetic energies in relation to the rotational energies.
In these clouds,  fragmentation only occurs in the high-density evolution phase ($n\gtrsim 10^{17}\cm$) after the clouds experience significant reduction of the magnetic field owing to Ohmic dissipation in the period of $10^{12}\cm \lesssim n \lesssim 10^{15}\cm$.
Fragments appearing in this phase have separations of $\lesssim 0.3$\,AU, and are expected to evolve into close binary systems. 
As a result, we found two typical fragmentation epochs, which cause different stellar separations.
Although these typical separations are disturbed in the subsequent gas accretion phase, we might be able to observe two peaks of binary separations in extremely young stellar groups.
\end{abstract}


\section{INTRODUCTION}
Stars are born as binary or multiple systems.
Observations have shown that, although the multiplicity depends on the stellar mass and surrounding environment, about 60-80\% of field stars are members of binary or multiple systems \citep{heintz69,abt76,abt83,duq91,fischer92}.
Thus, it has been established that a majority of main-sequence stars consist of binary or multiple systems.
Between multiples, the fraction of binary (b), triple (t), and higher-order multiple (h) systems is b: t: h = 0.75: 0.175: 0.04 \citep{duq91,tokovinin02}, which indicates that binary systems account for a large percentage of these systems \citep{goodwin07}. 
Observations of star-forming regions \citep[e.g.,][]{mathiu94} have shown that the multiplicity of pre-main sequence stars is larger than that of main-sequence stars.
For example, in the Taurus star-forming region, it is expected that almost 100\% of pre-main sequence stars are members of binary or multiple systems \citep{leinert93,kohler98}.
Recently, extremely young protostars (i.e., Class 0 protostars) have been observable with radio interferometer \citep{looney00,reipurth00} and wide-field near-infrared cameras \citep{haisch02,haisch04,duchene04}.
These observations have shown that stars already have a high multiplicity at the moment of their birth.
In contrast, \citet{lada06} pointed out that the multiplicity of low mass main-sequence stars ($\lesssim 0.5\msun$) is less than $\sim$ 50\%.
The multiplicity of the field stars tends to decrease with decreasing stellar mass.
\citet{duchene06} denoted that the multiple system decays by dynamical disruptive interaction in a short timescale ($\lesssim 1$Myr), and these systems evolve into an ejected single star or stable binary system.
This mechanism is effective in multiple systems composed of lower mass stars, because the binding energy of these systems is small.
Thus, it is considered that stars are born as binary or multiple systems, and some systems disrupt into  single stars as time goes on.
For this reason, binary and multiple frequencies decrease with time, especially in low-mass systems.

Distribution of binary separation is known to be very wide ($0.01$\,AU$ \lesssim r \lesssim 10^4$AU) and flat, usually modeled as a lognormal with a mean $\sim 30$\,AU \citep{goodwin07}.
For field main-sequence stars, roughly a third of all companions have close separation ($r\lesssim 1$AU). 
In addition, some binary systems have rotation periods of $P<10$\,days, which corresponds to $r\lesssim0.1$\,AU of the separation when the total mass of 1 $\msun$ is assumed \citep{mathiu94}, and they are classified as close binary
\footnote{
In this paper, for convenience, we define the binary system with separation of $r<0.1$\,AU as close binary, while the binary system with separation of  $r\ge0.1$\,AU as wide binary. 
}. 
These are the same for stars in the star-forming region.
\citet{mathiu94} showed that the binary separation extends from 0.01\,AU up to 1000\,AU also in pre-main sequence stars.
This indicates that both wide and close binary systems are formed in the star-formation process.
However, the mechanism that determines the binary frequency and separation is still unknown.
The frequency and separation distributions of binary are important clues to understanding the formation process of such stars.
It is, however, difficult to observe the formation process of binary systems directly, because these systems are embedded in dense gas clouds, and star (or binary) formation timescales are expected to be extremely short.
Thus, numerical calculations are necessary to understand the binary formation process including binary frequency and separation distributions.

Under the spherical symmetry, \citet{larson69}, \citet{tohline82}, and \citet{masunaga00} numerically investigated the evolution of a cloud from the molecular cloud core stage to that of protostar formation including the detailed radiative transfer.
They found that the gas cloud collapses self-similarly to reach stellar density.
According to the thermal evolution derived in their studies, cloud evolution is divided into four phases: the isothermal ($\nc\lesssim 10^{11}\cm$), adiabatic ($10^{11}\cm \lesssim \nc \lesssim 10^{16}\cm$), second collapse ($10^{16}\cm\lesssim \nc \lesssim 10^{21}\cm$), and protostar formation phases ($\nc \gtrsim 10^{21}\cm$), where $\nc$ means the number density at the center of the cloud.
The cloud collapses isothermally for $\nc \lesssim 10^{11}\cm$ (the isothermal phase or first collapse phase), then the gas around the center of the cloud becomes optically thick and collapses adiabatically after the central density exceeds $\nc \gtrsim 10^{11}\cm$ (the adiabatic phase).
In the adiabatic phase, the adiabatic core (or the first core) surrounded by a shock is formed.
When the central density reaches $\nc \simeq 10^{16}\cm$, the molecular hydrogen begins to dissociate and the cloud collapses rapidly (the second collapse phase).
After the molecular hydrogen is completely dissociated, the gas collapses adiabatically again and the protostar (or second core) is formed at $\nc \simeq 10^{21}\cm$.  
In these spherically symmetric calculations, however, cloud fragmentation or binary formation cannot be investigated, because these calculations only include the isotropic forces of the gravity and thermal pressure.
Cloud fragmentation is controlled by the anisotropic forces caused by the cloud rotation, magnetic field, and turbulence.
Thus, to investigate the fragmentation of the cloud, three-dimensional calculations are needed.

The evolution of rotating clouds in three dimensions have been investigated  by many authors \citep[c.f.,][]{bodenheimer00,goodwin07}.
\citet{miyama84}, and \citet{tsuribe99} calculated the evolutions of spherical clouds with initially uniform density and rigid-body rotation law in the isothermal regime.
They have shown that such clouds fragment if $\alpha_0\, \beta_0<0.12-0.15$, where $\alpha_0  \equiv E_{\rm T}/E_{\rm g}$ is the initial ratio of the thermal to the gravitational energy, and $\beta_0 \equiv E_{\rm R}/E_{\rm g} $ is the initial ratio of the rotational to the gravitational energy.
When fragmentation occurs in the isothermal phase,  the separation of the fragments has a scale of $100-10^4$\,AU, which corresponds to the Jeans length in this phase.
However, observations have shown that molecular clouds are centrally condensed, and often exhibit density profiles resembling those of the Bonnor-Ebert sphere \citep[e.g.,][]{alves01,kandori05}. 
\citet{boss93} calculated the evolution of such clouds with exponential density profiles in the isothermal regime, and he found that fragmentation occurs only in  highly unstable clouds of $\alpha_0 \lesssim 0.3$.
\citet{kandori05} found, from their near-infrared imaging observations, that many starless cores are near the stable state.
\citet{tachihara02} observed 179 cores in the star-forming regions, and showed that many of the cores are nearly in virial equilibrium.
Namely, their observations indicate that $\alpha_0$ is not so small, but is close to $\alpha_0\simeq1$
\footnote{
The critical Bonnor Ebert sphere has $\alpha_0 \simeq 0.7$ \citep{matsu03}.
}.
For this reason, it is expected that fragmentation rarely occurs in the isothermal phase.
The evolution of rotating clouds with centrally condensed density profiles from the isothermal to the adiabatic phase are investigated by \citet{matsu03} and \citet{cha03}.
They found that cloud rotation promotes fragmentation in the adiabatic phase.
\citet{matsu03} showed that fragmentation occurs in the adiabatic phase when the cloud core has a rotation energy of $\beta_0\gtrsim 3\times 10^{-3}$, which is sufficiently smaller than the observations \citep{goodman93,caselli02}.
\citet{caselli02} showed that the molecular clouds have $10^{-4}\lesssim \beta_0 \lesssim 0.02$ with the mean value of $\beta_0 \sim 0.02$.
Thus, their results indicate that the major fraction of molecular clouds might fragment in the adiabatic phase.

Many observations have shown that cloud cores have a significant non-thermal motion, which can be attributed to turbulence.
The turbulence is thought to dissipate as a cloud collapses \citep[e.g.,][]{larson81}.
In the very early phase of cloud collapse, however, the cloud acquires the angular momentum from turbulence, besides the large-scale ordered rotation motion of the host cloud.
Thus, turbulence promotes fragmentation, as for the cloud rotation.
Numerical studies have shown that more fragments appear in a cloud with larger turbulent energy in the initial state \citep{goodwin07}.
The cloud evolution with larger turbulent energy, in which an initial cloud has turbulent energy comparable to the gravitational, is investigated by \citet{bate02,bate03}, \citet{bate05}, \citet{delgado04a,delgado04b}.
In their calculations, fragmentation frequently occurs in an isolated cloud core, and many binary or multiple systems appear.
\citet{goodwin04a,goodwin04b} calculated the evolution of clouds with smaller turbulent energy.
They found fragmentation occurs and a binary system is born when the initial cloud has $\sim5\%$ turbulence energy exceeding gravitational energy.
They have suggested that almost all stars are born as binary or multiple systems, even if a cloud core has a low level of turbulence.
However, binary frequencies found in these turbulence calculations are higher than in observations.
Since the magnetic fields were ignored in these calculations, these high binary frequencies may be caused by neglect of the magnetic effect.

While cloud rotation and turbulence promote fragmentation, the magnetic field suppresses fragmentation.
The evolution of magnetized clouds from the isothermal to the adiabatic phase is investigated by \citet{hosking04}, \citet{machida04,machida05b}, \citet{ziegler05}, \citet{fromang06}, and \citet{price07}.
In their studies, fragmentation rarely occurs in strongly magnetized clouds.
This is because the angular momentum, which promotes fragmentation, is removed by the magnetic effect (e.g., magnetic braking, outflow, and jets).
\citet{machida05b} showed that fragmentation depends only on the ratio of  angular rotation speed to the magnetic field strength of the initial clouds, and hence it is necessary for strongly magnetized clouds to rotate rapidly for fragmentation.

Although fragmentation processes in the isothermal and adiabatic phase ($n\lesssim 10^{16}\cm$) are investigated by many authors, there are few studies of fragmentation in the second collapse and protostellar phases ($n \gtrsim 10^{16}\cm$).
When fragmentation occurs in the isothermal or adiabatic phases ($n \lesssim 10^{16}\cm$), fragments have separations of $10-10^4$\,AU, which correspond to the Jeans length in these phases.  
Fragments formed in these phases are expected to evolve into wide binary systems, when the binary systems maintain separation at their formation epoch. 
However, about 20-30\% of observed binaries have separations of $<10$\,AU \citep{mathiu94}.
In the collapsing cloud, the Jeans length becomes $\lesssim 10$\, AU after the central density exceeds $n\gtrsim 10^{13}\cm$.
Thus, fragments with the separation of $<10$\,AU and thus close binary systems are expected when fragmentation occurs for $n>10^{13}\cm$.
\citet{bate98} and \citet{whitehouse06} investigated the evolution of unmagnetized clouds from $10^6 \cm$ to $ 10^{21}\cm $.
According to their calculations, fragmentation does not occur in the high-density region, because the non-axisymmetric structure which appears in the adiabatic stage effectively removes angular momentum from the center of the cloud.
Thus, they concluded that fragmentation does not occur for $n\gtrsim 10^{16}\cm$.
However, they investigated the cloud evolution in a few parameters.
\citet{goodwin07} pointed out that cloud evolution in high-density gas is complicated, hence a large number of calculations and a statistical approach are needed to understand the fragmentation process.

\citet{banerjee06} investigated the cloud evolution of magnetized clouds for $10^6\cm \lesssim n \lesssim 10^{20}\cm$, and found that fragmentation occurs in the high density gas via a ring structure. 
They expected that this ring structure evolves into a close binary with the separation of $3\rsun$.
However, they adopted the ideal MHD approximation, which is not valid in high-density gas as  $n \gtrsim 10^{12}\cm$.
A significant magnetic flux is lost by Ohmic dissipation in the density range of $10^{12} \cm \lesssim n \lesssim 10^{15}\cm$ \citep{nakano02} and hence they overestimated the magnetic flux in a collapsing cloud, especially in the high-density gas regions.
In addition, they investigated the cloud evolution with only one parameter.

In this study, we investigate the evolutions of magnetized cloud cores ($\nc\simeq10^4 \cm$, $r=4.6\times 10^5$\,AU) until a protostar is formed ($\nc\simeq10^{22} \cm$, $r \sim 1 R_\odot$) using three-dimensional resistive MHD nested grid simulations.
We calculated 147 models with different magnetic field strengths, rotation speeds, and initial amplitudes of non-axisymmetric perturbation.
We found that the formation conditions for close and wide binary systems are related to the ratio of the rotational to the magnetic energy of the cloud core.

The structure of the paper is as follows: 
The framework of our models is given in \S 2 and in \S3 the numerical method of our computations is shown.
The numerical results are presented in \S 4.  
We discuss the fragmentation conditions in \S 5.

\section{MODEL}
 Our initial settings are almost the same as those of \citet{machida06a,machida06b,machida07a,machida07b}.
To study the cloud evolution, we use a three-dimensional resistive MHD nested grid code. 
We solve the resistive MHD equations including the self-gravity:  
\begin{eqnarray} 
& \dfrac{\partial \rho}{\partial t}  + \nabla \cdot (\rho \vect{v}) = 0, & \\
& \rho \dfrac{\partial \vect{v}}{\partial t} 
    + \rho(\vect{v} \cdot \nabla)\vect{v} =
    - \nabla P - \dfrac{1}{4 \pi} \vect{B} \times (\nabla \times \vect{B})
    - \rho \nabla \phi, & 
\label{eq:eom} \\ 
& \dfrac{\partial \vect{B}}{\partial t} = 
   \nabla \times (\vect{v} \times \vect{B}) + \eta \nabla^2 \vect{B}, & 
\label{eq:reg}\\
& \nabla^2 \phi = 4 \pi G \rho, &
\end{eqnarray}
 where $\rho$, $\vect{v}$, $P$, $\vect{B} $, $\eta$ and $\phi$ denote the density, 
velocity, pressure, magnetic flux density, resistivity, and gravitational potential, respectively. 
The resistivity $\eta$ in equation~(\ref{eq:reg}) is a function of the density and the temperature \citep{nakano02}. 
We use the resistivity $\eta$ as the value adopted in \citet{machida06b,machida07a}.
 To mimic the temperature evolution calculated by \citet{masunaga00}, we adopt a piece-wise polytropic equation of state as:
\begin{equation} 
P = \left\{
\begin{array}{ll}
 c_{\rm s,0}^2\, \rho & (\rho < \rho_c), \\
 c_{\rm s,0}^2\, \rho_c \left( \frac{\rho}{\rho_c}\right)^{7/5} & (\rho_c < \rho < \rho_d), \\
 c_{\rm s,0}^2\, \rho_c \left( \frac{\rho_d}{\rho_c}\right)^{7/5} \left( \frac{\rho}{\rho_d} \right)^{1.1}
 & (\rho_d < \rho < \rho_e), \\
 c_{\rm s,0}^2\, \rho_c \left( \frac{\rho_d}{\rho_c}\right)^{7/5} \left( \frac{\rho_e}{\rho_d} \right)^{1.1}
 \left( \frac{\rho}{\rho_e}   \right)^{5/3}
 & (\rho > \rho_e), 
\label{eq:eos}
\end{array}
\right.  
\end{equation}
 where $c_{\rm s,0} = 190$\,m\,s$^{-1}$, 
$ \rho_c = 3.84 \times 10^{-13} \, \rm{g} \, \cm$ ($n_c = 10^{11} \cm$), 
$ \rho_d = 3.84 \times 10^{-8} \, \rm{g} \, \cm$  ($n_d =  10^{16} \cm$), and
$ \rho_e = 3.84 \times 10^{-3} \, \rm{g} \, \cm$  ($n_e = 10^{21} \cm$).
For convenience, we define `the protostar formation epoch' as that at which the central density ($n_c$) reaches  $\nc = 10^{21} \cm$.
We also call the period of $n_c < 10^{11}\cm$ `the isothermal phase',  the period of $10^{11}\cm < \nc < 10^{16}\cm$ `the adiabatic phase', the period of $10^{16}\cm < \nc < 10^{21}\cm$ `the second collapse phase', and the period of $\nc > 10^{21}\cm$ `the protostellar phase.'

As the initial condition of the cores, we use the density profile increased by a factor $f$ (density enhancement factor) from the critical Bonnor-Ebert sphere \citep{ebert55, bonnor56}.
Note that the critical Bonnor-Ebert sphere has a density contrast of 14 between the center and the surface of the cloud.
The density enhancement factors are fixed as $f=1.68$ in all models except for five models shown in \S5.1.
In \S5.1, we adopt the density enhancement factors of $f$=1.4, 1.8, 2.8, 8.41, and 84.1.
The density enhancement factor is related to the stability of the initial cloud. 
We discuss the relation of the density enhancement factor and the stability of the cloud in \S5.1.
For the critical Bonnor-Ebert sphere, the central density is equal to $n_{\rm c,0}=10^4\cm$ ($\rho_0=3.82\times 10^{-20}$\,g$\cm$), and the initial temperature is 10K.
The radius of the critical Bonnor--Ebert sphere $R_c = 6.45\, c_s [4\pi G \rho_{BE}(0)]^{-1/2}$ corresponds to $ R_c = 4.58 \times 10^4$\,AU.
Outside this radius, we assume a uniform gas with a density of $n_{\rm BE}(R_{\rm c})=711\cm$.
The total mass contained in the critical Bonnor--Ebert sphere is $M_{\rm c} = 4.5\msun$, and our initial core is $f$ times as massive as this value.

Initially the cloud rotates rigidly $\Omega_0$ around the $z$-axis and has a uniform magnetic field $B_0$ parallel to the $z$-axis (or rotation axis).
The initial model is characterized by three nondimensional parameters: the magnetic-to-thermal pressure ratio,
\begin{equation}
\pb =  \dfrac{B_{\rm zc,0}^2}{4\pi \rho_{\rm c,0} c_{\rm s,0}^2},
\label{eq:alpha}
\end{equation}
the angular rotation velocity normalized by the free-fall timescale,
\begin{equation}
\omega = \dfrac{\Omega_{\rm c,0}}{ \sqrt{4 \pi  G\, \rho_{\rm c,0}} },
\label{eq:omega}
\end{equation}
and the initial amplitude of the non-axisymmetric perturbation $ \ap $.
We add $m=2$-mode non-axisymmetric density perturbation to the spherical initial cores.
Then, the density profile of the core is denoted as  
\begin{eqnarray}
\rho(r) = \left\{
\begin{array}{ll}
\rho_{\rm BE}(r) \, (1+\delta \rho)\,f & \mbox{for} \; \; r < R_{c}, \\
\rho_{\rm BE}(R_c)\, (1+\delta \rho)\,f & \mbox{for}\; \;  r \ge R_{c}, \\
\end{array}
\right. 
\end{eqnarray}
where $\rho_{\rm BE}(r)$ is the density distribution of the critical 
BE sphere, and $\delta \rho$ is the axisymmetric density perturbation. 
For the $m=2$-mode, we chose
\begin{equation}
\delta \rho = A_{\phi} (r/R_{\rm c})\, {\rm cos}\, 2\phi, 
\end{equation}
where $A_{\phi}$ represents the amplitude of the perturbation.

We adopt $\pb$ as $\pb=0-4$ as summarized in Table~\ref{table:alpha},  $\omega$ as $\omega=0.003-0.2$ as summarized in Table~\ref{table:omega}, and $\ap$ as $\ap=0.01, 0.2$, and 0.4.
The ratios of rotational ($\beta_0$) and magnetic ($\gamma_0$) energies to the gravitational energy
are summarized in each table.
In addition, we estimate the mass-to-flux ratio 
\begin{equation}
\dfrac{M}{\Phi} = \frac{M}{\pi R_{c}^2 B_0},
\end{equation}
where $M$ is the mass contained within the critical radius $R_{c}$ and  $\Phi$ is the magnetic flux threading the cloud.
There exists a critical value of $M/\Phi$ below which a cloud is supported against the gravity by the magnetic field.
For a cloud with uniform density,  \citet{mouschovias76} derived a critical mass-to-flux ratio
\begin{equation}
\left(\dfrac{M}{\Phi}\right)_{\rm cri} = \dfrac{\zeta}{3\pi}\left(\dfrac{5}{G}\right)^{1/2},
\end{equation}
where the constant $\zeta=0.53$ for uniform spheres ($\zeta=0.48$ by recent careful calculation of \citealt{tomisaka88a,tomisaka88b}).
For convenience, we use the mass-to-flux ratio normalized by the critical value as
\begin{equation}
\left( \dfrac{M}{\Phi}\right)_{\rm norm} \equiv \left(\dfrac{M}{\Phi}\right)/\left(\dfrac{M}{\Phi}\right)_{\rm cri}.
\label{eq:crit}
\end{equation}
The relation of parameter $\pb$ and the mass-to-flux ratio is also summarized in Table~\ref{table:alpha}.
We made 147 models by combining the above parameters.

\section{NUMERICAL METHOD}
We adopt the nested grid method  \citep[for detail, see][]{machida05a,machida06a} to obtain high spatial resolution near the center.
Each level of a rectangular grid has the same number of cells ($ = 128 \times 128 \times 8 $),  although the cell width $h(l)$ depends on the grid level $l$.
The cell width is reduced by a factor of 1/2 as the grid level increases by 1 ($l \rightarrow l+1$).
We assumed mirror symmetry with respect to $z$=0.
The highest level of grids changes dynamically.
We begin our calculations with four grid levels ($l=1,2,3$, and $4$).
Box size of the initial finest grid $l=4$ is chosen to be equal to $2 R_{\rm c}$, where $R_c$  denotes the radius of the critical Bonnor-Ebert sphere. 
The coarsest grid ($l=1$), therefore, has a box size equal to  $2^4\, R_{\rm c}$. 
A boundary condition is imposed at $r=2^4\, R_{\rm c}$, where the magnetic field and ambient gas rotate at an angular velocity of $\Omega_0$ (for detail, see \citealt{matsu04}).
A new finer grid is generated whenever the minimum local Jeans length 
$ \lambda _{\rm J} $ becomes smaller than $ 8\, h (l_{\rm max}) $, where $h$ is the cell width. 
The maximum level of grids is restricted to $l_{\rm max} = 30$, in which the maximum density up to $\nc=5\times10^{24}\cm$ can be calculated safely.
Note that since the central density stopped increasing at $\nc \lesssim 10^{22} \cm$ in every model, the grids of $l=29$, and 30 were not generated in our calculation.
Since the density is highest in the finest grid, the generation of a new grid ensures the Jeans condition of \citet{truelove97} with a margin of  safety factor of 2.
We adopted the hyperbolic divergence $\vect{B}$ cleaning method of \citet{dedner02}.

\section{RESULTS}

In this section, we describe the evolution of the magnetized clouds parameterized by the magnetic field strength $\pb$, rotation rate $\omega$, and initial amplitude of the non-axisymmetric perturbation $\ap$.
We compare the models with different  $\pb$ and $\omega$, but fixed  $\ap$ in the following subsections. 
In \S4.1, we show the evolutions for initial small non-axisymmetric perturbation $\ap=0.01$.
Then, the cloud evolutions for initial large non-axisymmetric perturbations are shown in \S4.2 ($\ap=0.2$) and 4.3 ($\ap=0.4$).

\subsection{Model with Small Non-Axisymmetric Perturbations}
In this subsection, we show the evolution of clouds for a small non-axisymmetric perturbation $\ap=0.01$ at the initial state.  
Figure~\ref{fig:1} shows the cloud structures around the center of the cloud for each model at the end of the calculation.
In this figure, all the models have the same $\ap=0.01$, but different $\pb$ and $\omega$.
Each panel is placed according to the parameters $\pb$ ($x$-axis) and $\omega$ ($y$-axis).
Models located in the upper-right region have strong magnetic fields and rapid rotations, while models located in the lower-left region have  weak magnetic fields and slow rotations in the initial state.
In the blue region, fragmentation occurs in the adiabatic (first core) phase ($10^{11}\cm \lesssim \nc \lesssim 10^{16}\cm$), while fragmentation occurs after the second collapse ($\nc > 10^{16}\cm$) in the red region.
On the other hand, fragmentation is never seen in the gray region.
For convenience, we call the models in which fragmentation occurs in the adiabatic phase ($10^{11}\cm < \nc < 10^{16}\cm$)  `first fragmentation models', the models in which fragmentation occurs after the second collapse ($\nc > 10^{16}\cm$) `second fragmentation models', and the models which never experience fragmentation `non-fragmentation models.'

In both the first and second fragmentation models, we stopped the calculations when the Jeans condition was violated outside the central deepest-level grid.
In these models, after the fragments escape from the finest grid, the Jeans condition is violated in the coarser grid.
In non-fragmentation models, we stopped the calculations in the protostellar phase ($\nc > 10^{21}\cm$) after we confirmed that fragmentation was unlikely to occur.
We had to stop the calculations in models near the lower-left  [($\pb$, $\omega$)=(0, 0.007), (0, 0.01)] and upper-right corners [($\pb$, $\omega$) = (4, 0.2)] (green region), before a protostar or fragmentation appeared.
In the models with ($\pb$, $\omega$)=(0, 0.007) and (0, 0.01), although we calculated the cloud evolutions for $\sim 2000$\,yr after the first cores were formed, the formed first cores were stable and showed neither signature of collapse nor fragmentation.
In addition, in these models, although we calculated the cloud evolutions for a long computational time of $\sim$ 1200\,CPU-hours, these clouds do not collapse any more after the first core is formed.
Note that it takes about 400\,CPU-hours to calculate the cloud evolution until the protostar is formed for a typical model. 
The first cores in the models in the lower-left corner seem long-lived as shown in \citet{saigo06}.
On the other hand, the cloud indicates oscillation around the initial state without any indication of collapse in model ($\pb$, $\omega$) = (4, 0.2).
Since this model has the strongest magnetic field and most rapid rotation, the cloud does not collapse to form the first core.

In the first fragmentation models (blue region), the cloud fragments into several pieces in the adiabatic phase ($10^{11}\cm \lesssim \nc \lesssim 10^{16}\cm$).
Typical separations between two fragments are $\sep \simeq 20-200$\,AU in these models.
In general, these models have high rotational energies but low magnetic energies at the initial state.
In the second fragmentation models, the cloud fragments after the second collapse phase ($\nc \gtrsim 10^{16}\cm$).
The separations between the fragments in these models are $\sep \sim 0.01-0.5$\,AU.
Compared with the first fragmentation models, the second fragmentation models have lower rotational energies and higher magnetic energies.
In non-fragmentation models (grey region), protostars are formed without fragmentation.
Compared with the fragmentation models, non-fragmentation models have lower rotational energies and higher magnetic energies in the initial state.
Figure~\ref{fig:1} shows that fragmentation occurs in the lower density (or large scale; 20-200\,AU) in the clouds having larger rotational energies and smaller magnetic energies, and fragmentation rarely occurs in clouds having smaller rotation energies or larger magnetic energies.
In the models shown in Figure~\ref{fig:1}, the clouds evolve keeping an axisymmetry because of the small non-axisymmetric perturbation.
Thus, fragmentation is induced via ring structure in many models, and the formed protostars have outer axisymmetric disks.

\subsubsection{Typical Model ($\ap=0.01$)}
\paragraph{First Core Formation}
To see the evolution of a molecular cloud into a protostar in more detail, we plot a number of snapshots for a typical model in Figure~\ref{fig:2}.
The model corresponds to ($\pb$, $\omega$, $\ap$) = (0.1, 0.05, 0.01).
As listed in Tables~1 and 2, this model has $\gamma_0= 9.59 \times 10^{-2}$ and $\beta_0 = 8.24 \times 10^{-3}$, respectively.
Figure~\ref{fig:2}{\it a} shows the initial density distribution along the $z=0$, $y=0$, and $x=0$ planes.
We adopted a Bonnor-Ebert density profile and  the initial cloud has a spherical structure with the central density of $\nc = 1.68\times 10^{4}\cm$.

Figure~\ref{fig:2}{\it b} shows the cloud structure when the central density reaches $\nc = 7.3\times 10^{14}\cm$ (adiabatic stage).
In this model, when the central density reaches $\nc \simeq 5.1 \times 10^{12}\cm$, a shock appears ($r\approx 2$\,AU) and a first core is formed at the center of the cloud
\footnote{
In this paper, we use `first core formation' when the core surrounded by a clear shock boundary is formed in the adiabatic phase.
}
.
The first core is represented by a thick red line in Figure~\ref{fig:2}{\it b}.
The first core has an oblate shape as shown in the middle and lower panels of Figure~\ref{fig:2}{\it b}.
Since the cloud collapses along the $z$-axis, the central region becomes oblate before the first core formation.
To evaluate the core shape, we define the oblateness as  $ \ob \, \equiv \, (h _l h _s) ^{1/2} / h _z $, where $ h _l $, $ h _s $, and $ h _z$ mean the major, minor, and $ z $-axis length derived from the moment of inertia for the high-density gas of $ \rho \, \ge \, 0.1  \rho _{\rm c}  $ according to Matsumoto \& Hanawa (1999).
Figure~\ref{fig:3}{\it a} shows the evolution of this oblateness.
In the isothermal phase, the oblateness continues to increase from the initial state and a disk-like structure is finally formed, because both the Lorentz and centrifugal forces make the cloud oblate.
The oblateness reaches $\ob \simeq 5$, and it reaches a peak when the central density reaches $\nc \simeq 10^{11}\cm$.
In the adiabatic phase, the oblateness begins to decrease because the central region collapses adiabatically.
The oblateness is $\ob \simeq 1.2$ when $\nc \sim 10^{13}\cm$.

After first core formation, outflow appears near the first core.
In the middle and lower panels of Figure~\ref{fig:2}{\it b}, it is shown that the gas is outflowing from the central region inside the thick orange lines.
After the gas is supported by the thermal pressure, the magnetic field lines are strongly twisted because the rotational timescale becomes shorter than the collapse timescale.
The outflow is driven due to the twisted magnetic field lines and rotation of the first core.
Outflows from the first cores are shown also in \citet{tomisaka98,tomisaka00,tomisaka02}, \citet{banerjee06}, \citet{matsu04}, \citet{fromang06}, and \citet{machida04,machida05b,machida07a,machida07b}.

Figure~\ref{fig:2}{\it c} shows the snapshot around the first core at the age of $263.4$\,yr after the first core formation.
Although the first core has an almost axisymmetric structure at its formation epoch (Fig.~\ref{fig:2}{\it b} upper panel), the non-axisymmetric perturbation grows and a spiral pattern appears inside the first core, as shown in the upper panel of Figure~\ref{fig:2}{\it c}.
To evaluate the degree of non-axisymmetric pattern, we define the axis ratio as  $ \ar \, \equiv \, h_l / h_s -1 $, according to Matsumoto \& Hanawa (1999).
Figure~\ref{fig:3}{\it b} shows the evolution of the axis ratio.
The axis ratio decreases for $10^{4}\cm \lesssim \nc \lesssim 10^{8}\cm$, because we start with a nearly equilibrium state \citep[for detail, see][]{machida05a, machida05b}.
Then, the axis ratio increases in proportion to $\ar \sim\rho^{1/6}$ for $10^8\cm \lesssim \nc \lesssim 10^{11}\cm$.
This growth rate corresponds to the result of the linear theory.
(The evolution of axis ratio can be described by the linear theory developed by \citealt{hanawa00} and \citealt{lai00}.)
In the linear regime, the degree of the deformation (i.e., the amplitude of the axis ratio) grows as $\ar \propto \rho^n$ with $n$ being a constant for a given polytropic index $\gamma$. 
For the isothermal gas ($\gamma = 1$), the power index of the density is  $n=1/6$ ($\ar \propto \rho^{1/6}$).
The power index of the density $n$ decreases with increasing polytropic index $\gamma$.
For $\gamma > 1.1$, the index $n$ becomes negative.
Namely, the axis ratio increases ($n>0$) for $\gamma < 1.1$, while the axis ratio decreases ($n<0$) for $\gamma>1.1$. 
The power index $n$ is given in Figure~14 of \citet{omukai05}. 
In the adiabatic phase, the axis ratio decreases for $10^{12}\cm \lesssim \nc \lesssim 10^{14}\cm$.
This is because the polytropic index $\gamma$ is changed from $\gamma=1$ to $\gamma=1.4$.

The axis ratio increases very rapidly near $\nc \simeq 3\times10^{14}\cm$, because collapse of the gas slows at this phase.
The non-axisymmetric perturbation can grow sufficiently in the quasi-static disk for the isothermal phase (\citealt{nakamura97,matsu99}, see also \citealt{machida07c}).
\citet{durisen86} shows that the bar structure develops with the bar mode instability, when the ratio of the rotational to gravitational energies ($\beta$) of the core exceeds $\beta>0.274$.
\citet{saigo07} found that the first core forms a bar or spiral pattern after exceeding $\beta \gtrsim 0.27$.
Figure~\ref{fig:3}{\it c} shows the evolution of angular velocity normalized by the square root of the density [$\Omega_0/(4 \pi G \rho)^{1/2} \equiv \omega_{\rm c}$: the normalized angular velocity].
This panel shows that the normalized angular velocity saturates around $\omega_{\rm c} \simeq 0.1-0.2$ in the isothermal phase, while it increases in the adiabatic phase.
After the first core is formed, the normalized angular velocity oscillates intensely around $\omega_{\rm c}\simeq 0.2-0.6$.
The evolution of angular velocity in the collapsing cloud is described by \citet{machida05a,machida06a,machida07a}.
Since $\beta$ is related to $\omega_{\rm c}$ by $\beta=\omega_{\rm c}^2$ for a spherical cloud with a constant density, the first core is expected to have $\beta=0.04-0.36$.
Thus the bar mode instability can be induced in the first core by the dynamical instability \citep{durisen86}.
As a result, the first core shown in Figure~\ref{fig:2}{\it c} upper panel has a spiral structure, which is thought to be caused by nonmagnetic instability.

Figure~\ref{fig:2}{\it c} upper panel shows two density peaks near the center of cloud, which are caused by the fragmentation of the first core.
In this model, fragmentation and subsequent merger are repeated twice in the range of $10^{15}\cm \lesssim \nc \lesssim 10^{18}\cm$.
The first fragmentation occurs at the density $\nc \simeq 2\times 10^{15}\cm$, and then, the fragments are merged in a short timescale.
The second fragmentation occurs at $\nc \simeq 8\times 10^{16}\cm$, and then the density exceeds the H$_2$ dissociation density and the second collapse occurs in each  fragment.
As shown in Figure~\ref{fig:3}{\it c}, the normalized angular velocity decrease rapidly in the range of $10^{16}\cm \lesssim \nc \lesssim 10^{18}\cm$.
This is driven by the angular momentum redistribution into orbital and spin motions.
The two fragments are merged to form a nearly spherical core at the density $\nc \simeq 2\times 10^{18}\cm$.
The normalized angular velocity increases again after reaching $\nc \simeq 2\times 10^{18}\cm$, because the fragments bring the angular momentum into the center of the cloud again.
After merger, the central region changes shape from spherical to bar-like.
At the protostar formation epoch ($\nc = 10^{21}\cm$), the central region has an axis ratio of $\ar = 2.8$ and a bar-like protostar is expected in this model.

\paragraph{Protostar Formation}
Figures~\ref{fig:2}{\it d}-{\it f} show the cloud evolution after the protostar is formed.
The axis ratio continues to increase in the protostellar phase, and an elongated bar is formed around the center of the cloud.
When the axis ratio reaches $\ar = 10.2$ at $\nc \simeq 1.8 \times 10^{21}\cm$, the bar fragments into three pieces  as shown in Figure~\ref{fig:2}{\it d}.
When the fragmentation occurs, the separation between fragments is equal to $\sep \simeq 0.031$\,AU.
After fragmentation, fragments approach each other, and they are merged into a single core as shown in Figure~\ref{fig:2}{\it e}.
After the merger, fragmentation occurs again through a ring-like structure as shown in Figure~\ref{fig:2}{\it f}.
The final fragments survive  separately, until we stop the calculation.
At the end of the calculation, the separation between fragments is equal to $\sep\simeq 0.03$\,AU, and each fragment has a mass of $M=3.1\times 10^{-3}\msun$.

\paragraph{Magnetic Field}
Figure~\ref{fig:3}{\it d}  shows the evolution of the magnetic field strength normalized by the square root of the central density [$B_{\rm zc}/(4 \pi c_{s,0}^2 \rhoc)^{1/2}$: the normalized magnetic field strength].
Since, in \citet{machida05a,machida06a,machida07a}, we already discussed in detail the evolution of the magnetic field in the collapsing cloud, in this paper we briefly show the evolution of the magnetic field.
In the isothermal phase, the normalized magnetic field strength is saturated at a constant level according to the magnetic flux-spin relation of \citet{machida05a,machida06a}, then the magnetic fields are dissipated by Ohmic dissipation in the range of $10^{12}\cm \lesssim \nc \lesssim 10^{15}\cm$, as shown in \citet{nakano02} and \citet{machida07a}.
The rapid decrease in Figure~\ref{fig:3}{\it d} for $10^{12}\cm\lesssim \nc \lesssim 10^{17}\cm$ corresponds to this Ohmic dissipation. 
The first core, which is traced by two spiral arms occupying $r\lesssim 3$\,AU, has a large beta $\beta\simeq 10-10^3$, because the gas density exceeds $\nc>10^{12}\cm$.
Thus, in the adiabatic phase, the magnetic field barely affects the cloud evolution inside the first core.
On the other hand, the region outside the spiral arm contains a strong magnetic field of $\beta\simeq 1$, because the gas is well coupled with the magnetic field for  $\nc \lesssim 10^{12}\cm$.
In the second collapse phase, since the Ohmic dissipation becomes ineffective for the gas in which the ionization degree is recovered by the electron for alkali metals, the magnetic field can be amplified again, as shown in Figure~\ref{fig:3}{\it d}.
At the end of the calculation, each protostar has a magnetic field of 1.08kG at its center.

\subsection{Model with Large Non-Axisymmetric Perturbations}
In this section, we show evolutions of clouds with large non-axisymmetric perturbations ($\ap=0.2$ and 0.4) at the initial state.

\subsubsection{Models with $\ap=0.2$}
Figure~\ref{fig:4} shows cloud structures around the center of the cloud at the end of calculations for models with $\ap=0.2$.
In the upper-left region (weak magnetic field and fast rotation), fragmentation occurs in the adiabatic phase, and two fragments have wide separations ($20 \lesssim R_{\rm sep} \lesssim 100$\,AU).
On the other hand, fragmentation occurs in the second collapse or after the protostellar phase with narrow separations ($R_{\rm sep} \lesssim 0.4$\,AU) in the lower region (red region).
Clouds having larger magnetic energies and smaller rotational energies at the initial state lead to formation of a protostar without fragmentation (grey region).
Thus, similarly to  models with small non-axisymmetric perturbation $\ap=0.01$, the magnetic field suppresses fragmentation, but the rotation promotes it also in the models with large non-axisymmetric perturbation $\ap=0.2$.
However, three regions of first (blue region), second (red region), and non-fragmentation (grey region) are clearly divided in Figure~\ref{fig:1}, while the boundary between these three regions is complicated in Figure~\ref{fig:4}.
Figures~\ref{fig:1} and \ref{fig:4} show that, even though the initial clouds have the same magnetic field strength and rotation rate, fragmentation epoch depends on the initial amplitude of the non-axisymmetric perturbation $\ap$.
For models of ($\pb$, $\omega$) = (0, 0.07), (0, 0.05), (0.001, 0.05), and (0.01, 0.05), fragmentation occurs in the adiabatic phase in the models with $\ap=0.01$ (Fig.~\ref{fig:1}), while fragmentation occurs in the second collapse or protostar phases in the models with $\ap=0.2$ (Fig.~\ref{fig:4}).
The epoch of fragmentation is delayed with increasing perturbation amplitude $\ap$.
In addition, some models in which fragmentation is found in Figure~\ref{fig:1} (models with $\ap=0.01$) do not fragment when the initial cloud has a large non-axisymmetric perturbation $\ap=0.2$.
For  models of ($\pb$, $\omega$) = (1, 0.03), (1, 0.05), (1, 0.07), and (1, 0.1), fragmentation occurs in the models with $\ap=0.01$ (Fig.~\ref{fig:1}), while fragmentation does not occur in the models with $\ap=0.2$ (Fig.~\ref{fig:4}).

As shown in \citet{machida05b, machida07b}, there are two modes of fragmentation: ring and bar fragmentations.
The fragmentation mode is related to the axis ratio and rotation rate of the central core.
When the cloud has a small axis ratio and rapid rotation, fragmentation is induced via a ring-like structure.
On the other hand, when the cloud has a large axis ratio ($\ar \gtrsim 10$), fragmentation occurs via a bar-like structure, irrespective of the cloud rotation rate. 
Non-axisymmetric structure has two distinct effects on fragmentation: suppression  and  promotion of fragmentations.
When a core (the first core or second core) has a moderate axis ratio ($1\lesssim \ar \lesssim 10$), the angular momentum is removed from the core by gravitational torque due to the non-axisymmetric pattern and thus fragmentation is suppressed.
In contrast, when the core has a large axis ratio ($\ar \gtrsim 10$), fragmentation occurs through the bar instability, irrespective of the cloud rotation rate.
Thus, compared with models having small non-axisymmetric perturbation $\ap=0.01$ (Fig.~\ref{fig:1}), in models with large non-axisymmetric perturbation $\ap=0.2$ (Fig.~\ref{fig:4}), fragmentation is suppressed in some models [e.g., models ($\pb$, $\omega$) = (0.001, 0.05), and  (0.01, 0.05 )] by removal of angular momentum owing to the gravitational torque, although fragmentation is promoted in some models by the prominent bar mode [($\pb$, $\omega$) = (0.001, 0.03)].
However, comparison between Figures~\ref{fig:1} and \ref{fig:4} indicates that only the models located near the boundaries between these three regions (first, second, and non-fragmentation) are influenced by the initial amplitude of the non-axisymmetric perturbation.
The trend that the magnetic field suppresses fragmentation while rotation promotes it is the same for both the models with $\ap=0.01$ and $\ap=0.2$.

\subsubsection{Typical Model ($\ap=0.2$)}
Figure~\ref{fig:5} shows the cloud structure for the model ($\pb$, $\omega$, $\ap$) = (0.01, 0.05, 0.2).
In this figure, the structures around the center of the cloud with different scales are plotted for the final state of the calculation.
The spatial scale of each successive panel differs by a factor of four and thus the scale between Figures~\ref{fig:5}{\it a} and \ref{fig:5}{\it f} differs by a factor of 1024.
In this model, fragmentation occurs in the second collapse phase, while fragmentation occurs in the adiabatic phase for the model with the same $b$ and $\omega$ but smaller $\ap=0.01$, as shown in Figure~\ref{fig:1}.
This is because the angular momentum in the model with $\ap = 0.2$ is more effectively transferred from the center of the cloud than in the model with $\ap=0.01$.
This stabilizes the first core for fragmentation.

Middle and lower panels of Figures~\ref{fig:5}{\it a} and {\it b} show a thin disk, which is formed in the adiabatic phase. 
Red-dotted lines in Figures~\ref{fig:5}{\it a} and {\it b} indicate a shocked region which corresponds to the first core.
Since this model has a large amount of non-axisymmetric perturbation $\ap=0.2$ in the initial state, the non-axisymmetric pattern grows sufficiently and a spiral appears just after the first core formation.
After that, the central compact core is separated from the ambient ring-like structure.
The upper panel of Figure~\ref{fig:5}{\it b} shows that the first core is composed of a central compact core and a surrounding ellipsoidal structure.
There is a clear gap between the central core and ambient medium.
In the middle and lower panels of Figures~\ref{fig:5}{\it a} and {\it b}, orange lines indicate the outflow.
Inside these lines, the gas is outflowing from the center of the cloud  with $\sim1.2\km$. 
The outflow appears when the central density reaches $2.1\times 10^{14}\cm$ in this model.

Figures~\ref{fig:5}{\it c} and {\it d} indicate that the central region has entered the second collapse phase because the density exceeds $\nc>10^{16}\cm$, while the surrounding ring-like structure stays in the adiabatic phase.
Although we cannot see any internal structure in the compact core in Figures~\ref{fig:5}{\it a} and {\it b}, Figures~\ref{fig:5}{\it c} and {\it d} show the central region forms a spiral structure.
Similar structure is seen in \citet{bate98}, who calculated the evolution of an unmagnetized cloud core having the parameter of $\omega=0.07$ with an initial number density of $n_{\rm c,0} = 3.6\times 10^5\cm$.
He found that the cloud collapses to form a single star without fragmentation after the second collapse phase, because the angular momentum is effectively removed from the center of the cloud by the spiral patterns.
Although our model contains the magnetic field,  the magnetic field becomes extremely weak in the second collapse phase due to the Ohmic dissipation, as shown in \S4.1.1 and \citet{machida07a}, thus, the magnetic field barely affects the cloud evolution in this phase.
For this reason, both the unmagnetized and magnetized cloud are thought to follows a similar evolutional pattern in the second collapse phase.

In the upper panels of Figures~\ref{fig:5}{\it e} and {\it f},  four fragments appear inside the spiral structure.
In this model, the first fragmentation occurs when the central density reaches $\nc=1.1\times 10^{18}\cm$, and two clumps appear.
They are located at ($x$, $y$) $\simeq$ ($\pm$0.01\,AU, $\pm$0.07\,AU) as shown in the upper panel of Figure~\ref{fig:5}{\it f}.
After that, fragmentation occurs again in each clump at $\nc \simeq 4\times 10^{21}\cm$, and as a result four fragments are formed as shown in Figure~\ref{fig:5}{\it f}.
Since the gas density of these four fragments exceeds $\nc \gtrsim 10^{21}\cm$, these fragments are protostars in our definition.
Thus, we found a quadruple protostellar system.
At the end of the calculation, the protostars have the furthermost separation  of $R_{\rm sep}\sim0.15$\,AU.
To investigate whether these protostars actually evolve into a quadruple stellar system without merger, further long calculation is needed.

In the protostellar phase, jet appeared from the center of the cloud with the velocity of $\simeq 18\km$.
The boundary between the inflow and jet is represented by orange lines in the upper and middle panels of Figures~\ref{fig:5}{\it d} and {\it f}, inside which the gas is outflowing from the protostar.
In this model, two flows appear: a low velocity flow is driven by the rotation of the  first core, and a high velocity flow is driven by the spinning motion of the protostar.
Since we investigated in detail the evolution and mechanism of outflow and jet in a separate paper \citep{machida07b}, we do not describe these further here.
The outflow and jet driven from the binary system will be investigated in a subsequent paper.

\subsubsection{Models with $\ap=0.4$}
\begin{figure}
\begin{center}
\includegraphics[width=160mm]{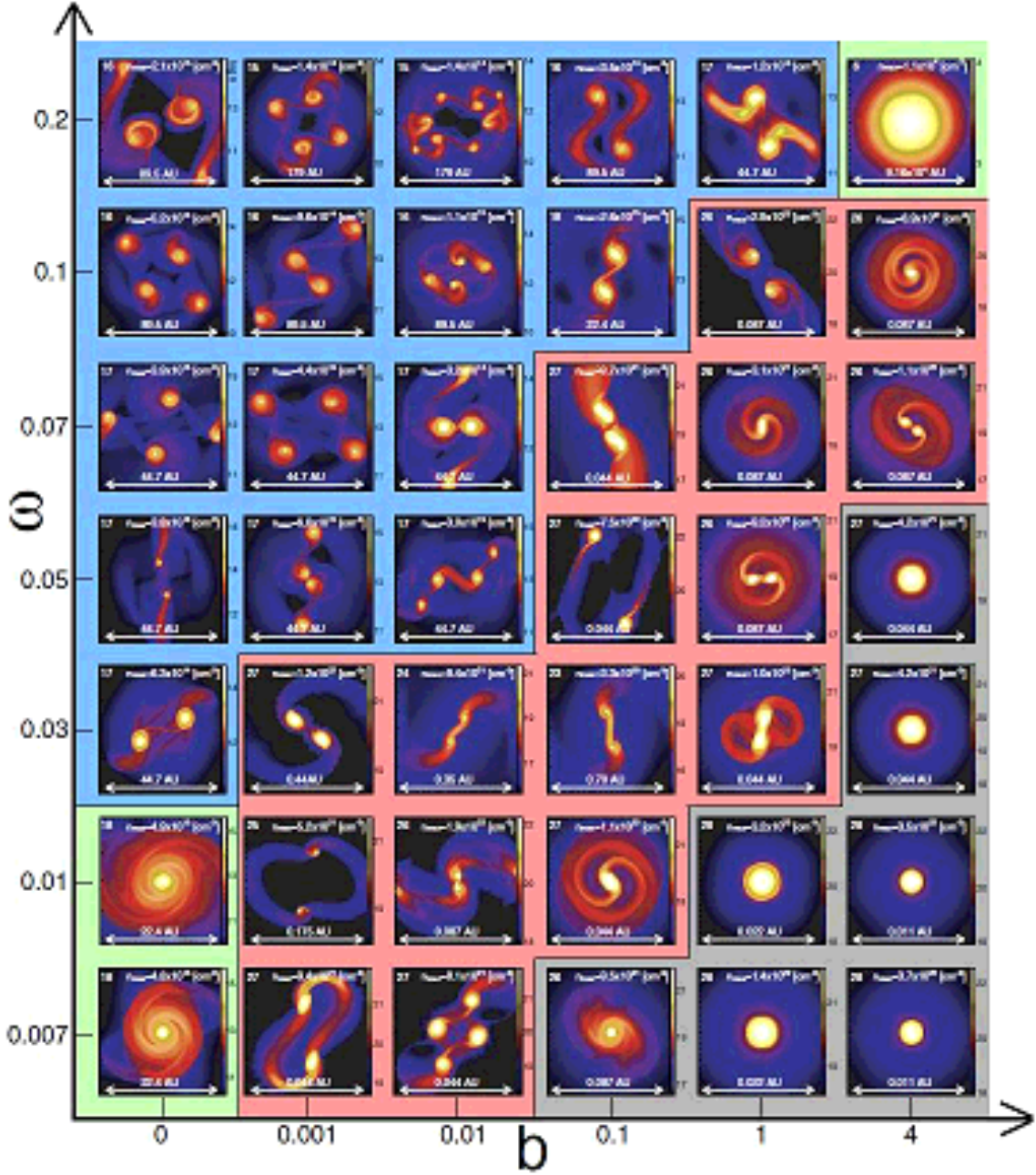}
\caption{
Final states for models with $\ap=0.01$ against parameters $\pb$, and $\omega$.
Density (color-scale) on the cross-section of the z=0 plane are plotted in each panel.
The grid level, maximum number density ($n_{\rm max}$), and grid scale are denoted inside each panel.
Background colors indicate that {\it Blue}: fragmentation occurs in the adiabatic phase, {\it Red}: fragmentation occurs only in the second collapse or protostellar phase, {\it Gray}: no fragmentation occurs through all phases of the cloud evolution, and {\it Green}: the cloud no longer collapses.
}
\label{fig:1}
\end{center}
\end{figure}

\begin{figure}
\begin{center}
\includegraphics[width=130mm]{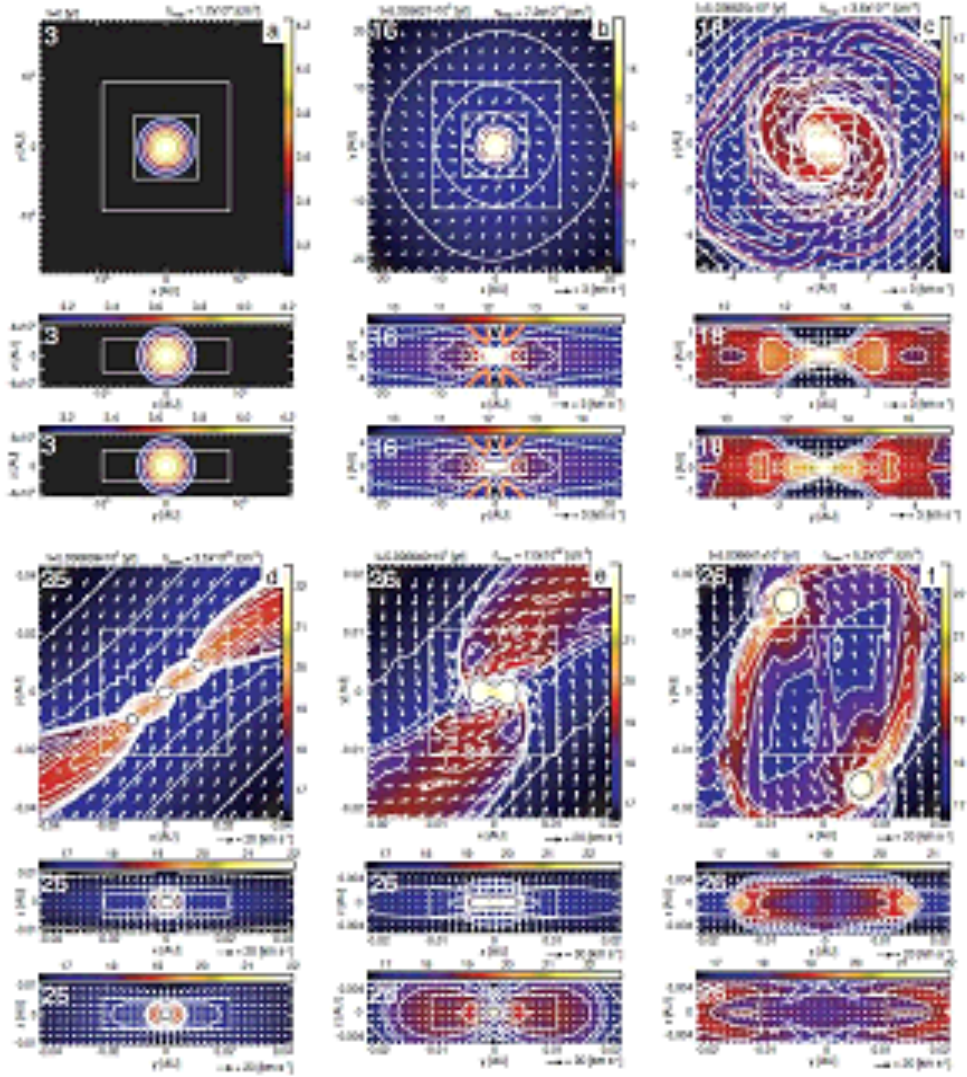}
\caption{
Cloud evolution for model ($\pb$, $\omega$, $\ap$) = (0.1, 0.05, 0.01).
Density (color-scale) and velocity distribution (arrows) on the cross-section of the $z=0$ plane (upper panels), $y=0$ plane (middle panels), and  $x=0$ plane (lower panels) are plotted.
Panels (a) through (f) are snapshots at the stages 
(a) $n_c =1.7 \times 10^4 \cm$ ($l=3-5$; initial state), \ 
(b) $ 7.3 \times 10^{14} \cm$ ($l=16$--19; adiabatic phase), \ 
(c) $ 3.6 \times 10^{17} \cm$ ($l=18$--20; second collapse phase), \
(d) $ 3.5 \times 10^{22} \cm$ ($l=25$, 26; protostellar phase), \
(e) $ 7.0 \times 10^{22} \cm$ ($l=26$, 27; protostellar phase), \ and \
(f) $ 5.2 \times 10^{22} \cm$ ($l=26$, 27; calculation end),
where $l$ denotes the level of the subgrid.
The red lines in panels {\it b} and {\it c} indicate the first core.
The black lines in panels {\it d}--{\it f} indicate the protostar (or second core).
The thick orange lines in panel {\it b} indicate the border between the infalling and outflowing gases (contour of $v_{ z} = 0$).
The level of the subgrid is shown in the upper left corner of each upper panel.
The elapsed time $t$, maximum number density $n_{\rm max}$, and arrow scale are denoted in each panel.
}
\label{fig:2}
\end{center}
\end{figure}

\begin{figure}
\begin{center}
\includegraphics[width=160mm]{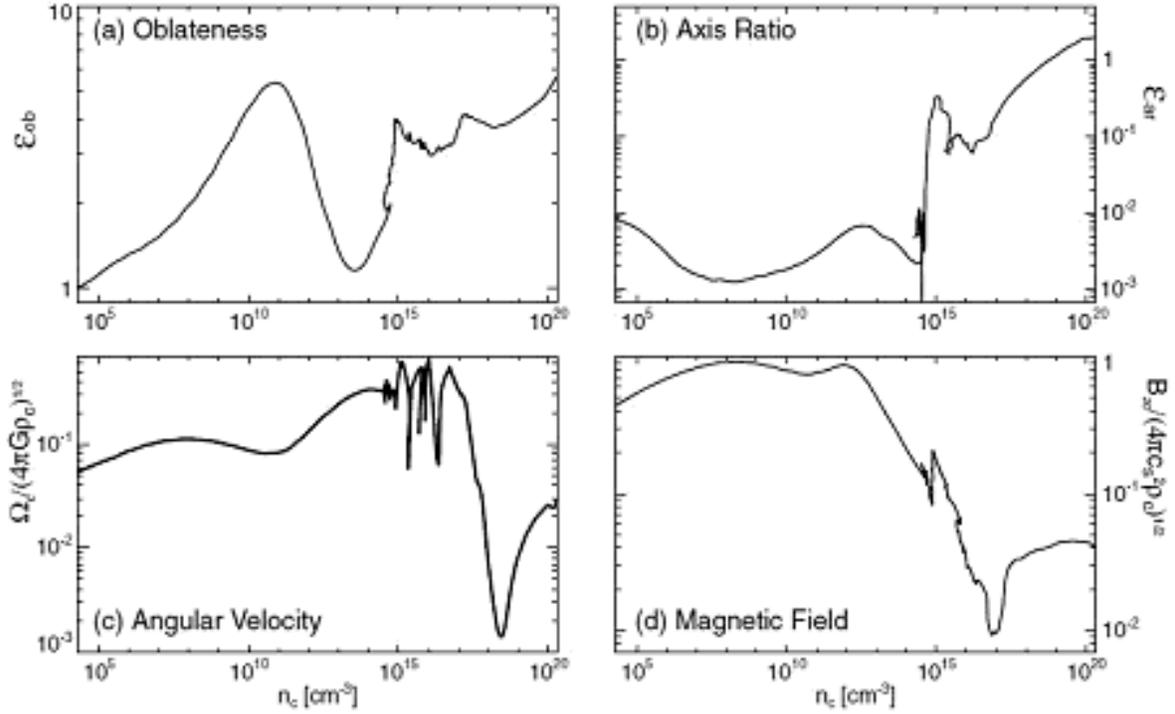}
\caption{
Evolution of ({\it a}) oblateness, ({\it b}) axis ratio, ({\it c}) angular velocity normalized by the square root of the central density $\Omega_{\rm c}/(4\pi G \rhoc)^{1/2}$, and ({\it d}) magnetic field strength normalized by the square root of the central density $B_{\rm zc}/(4\pi c_{\rm s,0}^2 \rhoc)^{1/2}$.
All quantities are plotted against the central density $\nc$ for model ($\pb$, $\omega$, $\ap$) = (0.1, 0.05, 0.01).
}
\label{fig:3}
\end{center}
\end{figure}

\begin{figure}
\begin{center}
\includegraphics[width=160mm]{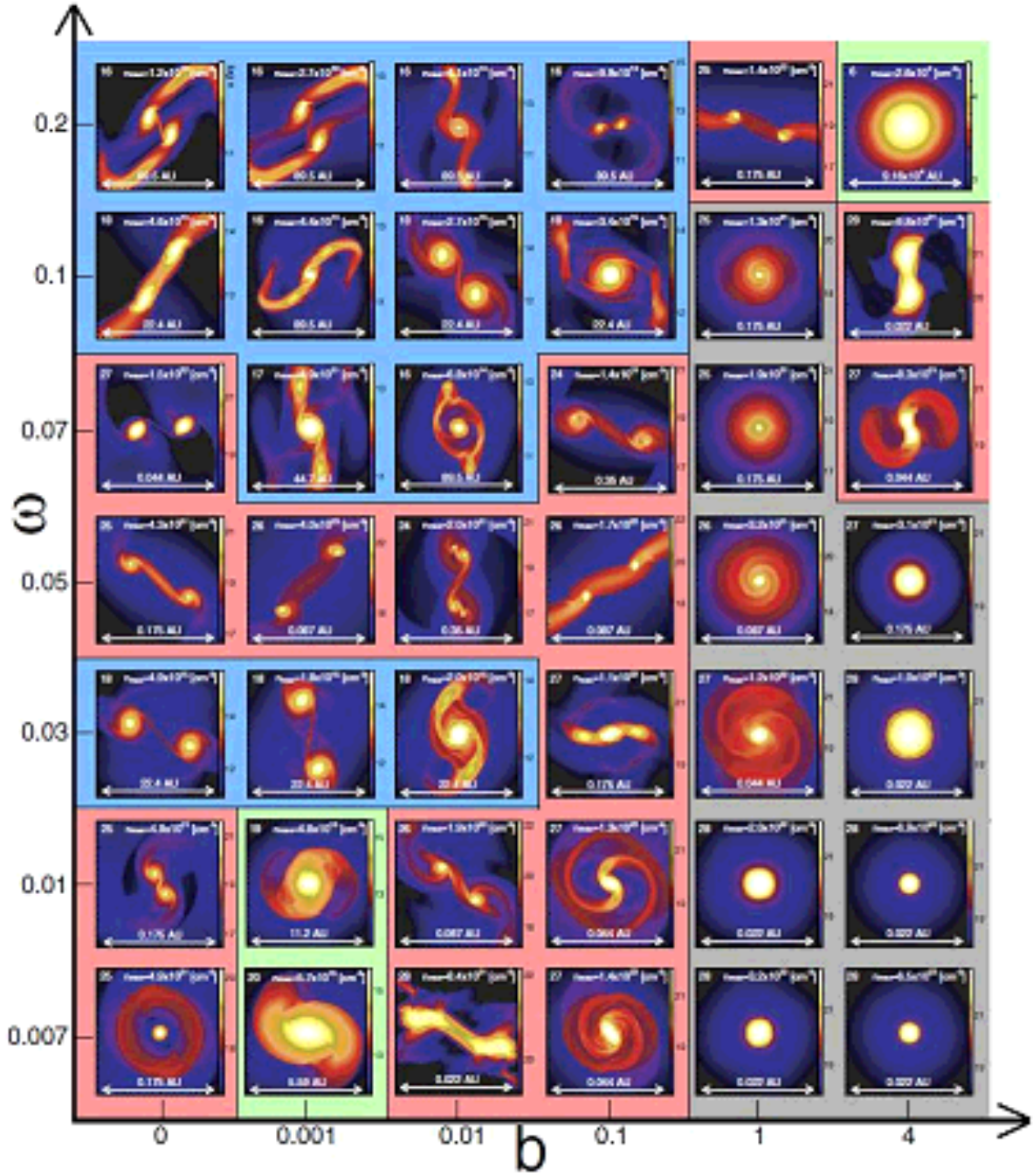}
\caption{
Same as Fig.~\ref{fig:1}, but for models with $\ap=0.2$.
}
\label{fig:4}
\end{center}
\end{figure}

\begin{figure}
\begin{center}
\includegraphics[width=160mm]{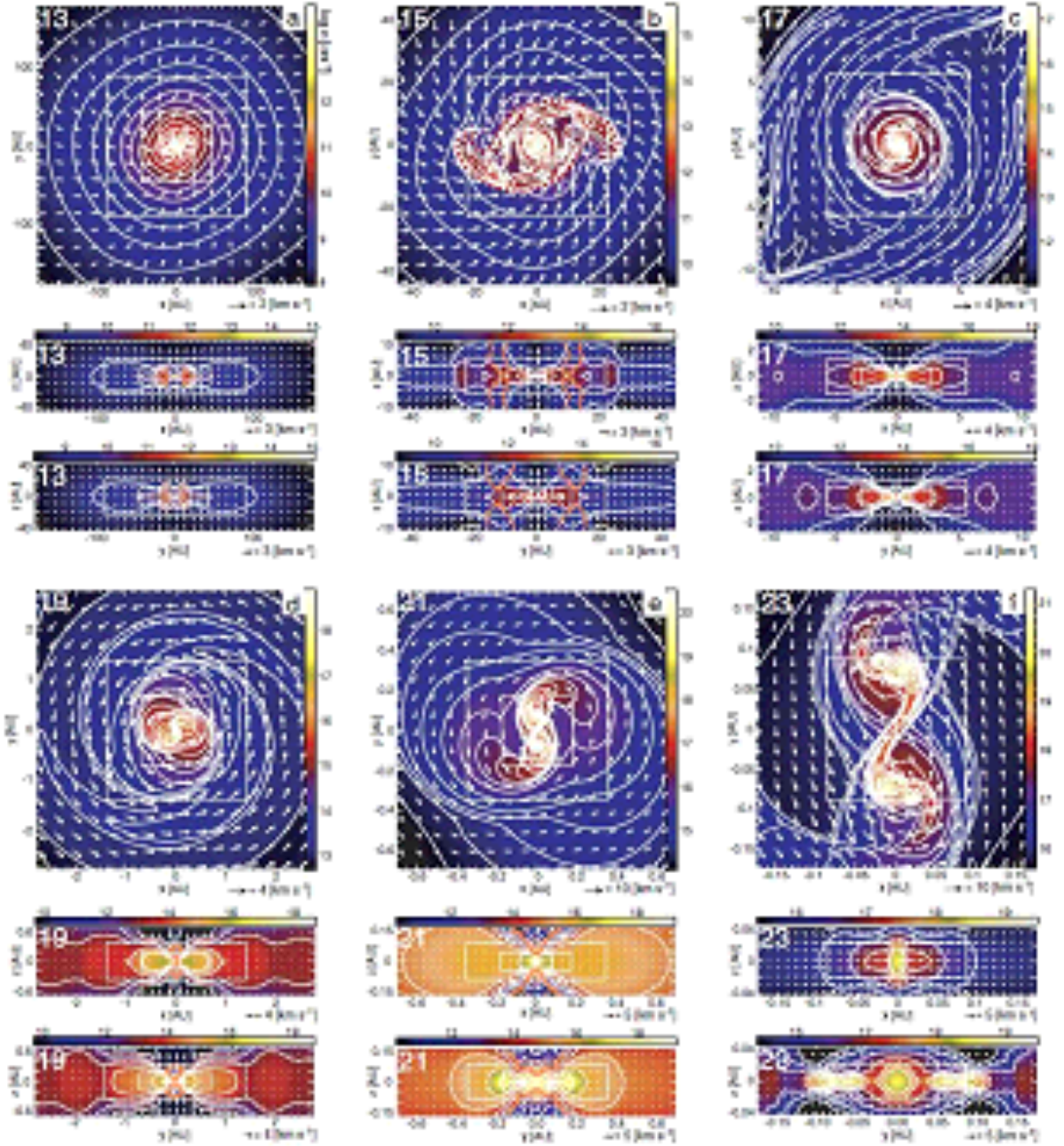}
\caption{
Cloud structure for model ($\pb$, $\omega$, $\ap$) = (0.01, 0.05, 0.2) at the end of the calculation with different scales.
Density (color-scale) and velocity distribution (arrows) on the cross-section of the $z=0$ plane (upper panels), $y=0$ plane (middle panels), and  $x=0$ plane (lower panels) are plotted.
The red dotted line in panel {\it b} indicates the first core.
The thick orange lines in panel {\it b} indicate the borders between the infalling and outflowing gases (contour of $v_{ z} = 0$).
The level of the subgrid is shown in the upper left corner of each upper panel.
Arrow scale is also denoted in each panel.
}
\label{fig:5}
\end{center}
\end{figure}

\begin{figure}
\begin{center}
\includegraphics[width=160mm]{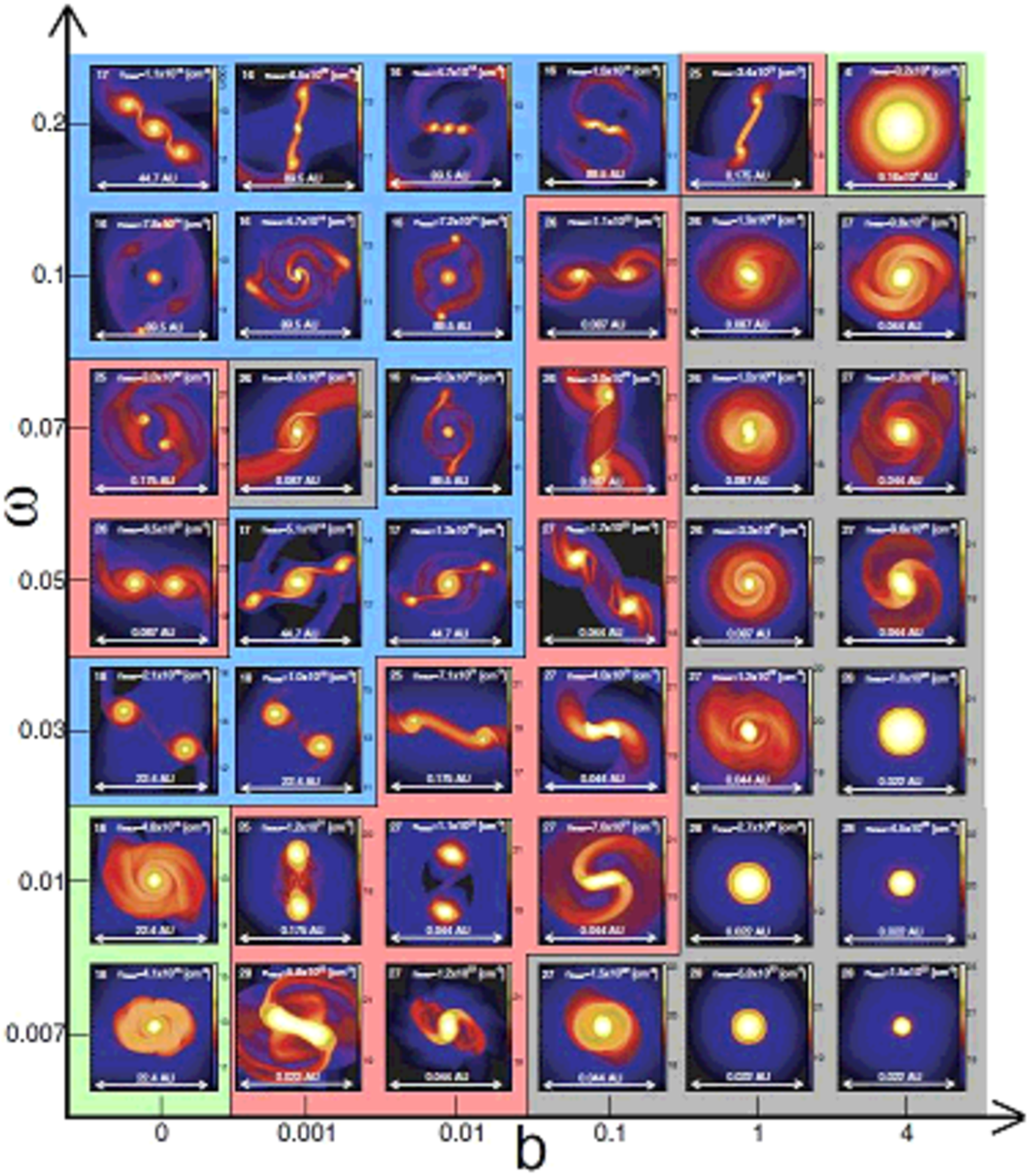}
\caption{
Same as Fig.~\ref{fig:1}, but for models with $\ap=0.4$.
}
\label{fig:6}
\end{center}
\end{figure}

Figure~\ref{fig:6} shows the cloud structures around the center of the cloud at the end of the calculation for models with $\ap=0.4$.
The fragmentation properties (fragmentation epoch, scale, number of fragments, and structure) shown in this figure are almost the same as those in Figures~\ref{fig:1} and \ref{fig:4}.
In the following, we only describe the differences between the models $\ap=0.4$ (Fig.~\ref{fig:6}), $\ap=0.01$ (Fig.~\ref{fig:1}) and $\ap=0.2$ (Fig.\ref{fig:4}).

Firstly, we compare the models having the same parameters of  ($\pb$, $\omega$) = (0.001, 0.07), which are located in the third row and the second column in Figures~\ref{fig:1}, \ref{fig:4}, and \ref{fig:6}.
Fragmentation occurs in the adiabatic phase with models having $\ap=0.01$ and $0.2$, while no fragmentation occurs through any of the phases of cloud evolution in the model with $\ap=0.4$.
In Figure~\ref{fig:6}, for this model, we can see that the central region around the protostar indicates a bar-like structure in a large scale, and a spiral structure in a small scale.
\citet{bate98} showed that the spiral structure effectively removes the angular momentum from the center of the cloud, and a single protostar is directly formed without fragmentation.
Also in this model,  since the angular momentum is effectively removed owing to the bar or spiral structure, the protostar is thought to form without fragmentation.

Next, we describe the models with ($\pb$, $\omega$) = (0.001, 0.05), and (0.01, 0.05), which are located at the fourth row and the second and third columns.
Figures~\ref{fig:1}, \ref{fig:4}, and \ref{fig:6} show that fragmentation occurs in the adiabatic phase with wide separations ($\sim 40$\,AU) for the models with $\ap=0.01$ and $0.4$,  while fragmentation occurs with a narrow separation ($\sim0.1$\,AU) in the second collapse or protostellar phase for the models with $\ap=0.2$.
According to the degree of non-axisymmetric pattern (i.e., the axis ratio), these models experience three different types of evolutions:  (i) if the core has a small axis ratio, fragmentation occurs via a ring-like structure in the adiabatic phase owing to the rapid rotation ($\ap=0.01$), (ii) if the core has a moderate axis ratio, fragmentation occurs in the second collapse phase, because the angular momentum is effectively removed by the bar-like structure in the adiabatic phase ($\ap=0.2$), and (iii) if the core has a large axis ratio, fragmentation occurs via a bar-like structure in the adiabatic phase, because a considerably elongated bar is formed in the adiabatic phase and such a structure is  unstable ($\ap=0.4$).
Therefore, the separation between fragments does not represent the initial amplitude of the nonaxisymetric perturbations.

Thirdly, we compare the models having the same ($\pb$, $\omega$) = (0.001, 0.03), which are located in the fifth row and the second column in Figures~\ref{fig:1}, \ref{fig:4}, and \ref{fig:6}.
In these models, fragmentation occurs in the second collapse phase for the model with $\ap=0.01$, while fragmentation occurs in the adiabatic phase for the models with $\ap=0.2$ and 0.4.
This difference is understood as follows: For the model with a small non-axisymmetric perturbation ($\ap=0.01$), since both the rotation rate and the non-axisymmetric perturbation are small when the first core is formed,  neither ring nor bar fragmentation occurs.
On the other hand, fragmentation occurs in the bar for models with larger non-axisymmetric perturbations ($\ap=0.2$ and 0.4), because a considerably elongated bar is formed at the first core formation epoch.

Finally, the strong magnetized models  $\pb\ge1$ are described.
Figures~\ref{fig:1}, \ref{fig:4}, and \ref{fig:6} clearly show that fragmentation becomes harder with increasing $\ap$ especially in the strong magnetized models.
For models with $\ap=0.01$ (Fig.~\ref{fig:1}), fragmentation occurs in the second collapse phase for models having $\omega=$ 0.03, 0.05, 0.07 and 0.1 for $\pb=1$, and $\omega$=0.07 and 0.1 for $\pb$=4.
In contrast, for models with $\ap=0.4$ (Fig.~\ref{fig:4}), no fragmentation occurs in models having the same $\pb$ and $\omega$.
The increase of the angular velocity is suppressed by the magnetic effect \citep[e.g., magnetic braking and outflow;][]{machida07a} in strongly magnetized models, and then these models manage to fragment in the second collapse or protostellar phase when the initial cloud has a small amount of the non-axisymmetric perturbation of $\ap=0.01$.
In contrast, for a models with large $\ap$, since the angular momentum is removed not only by the magnetic effect  but also the gravitational torque  as shown in Figure~\ref{fig:5}{\it c}-{\it f}, fragmentation is more severely suppressed.

Figures~\ref{fig:1}, \ref{fig:4}, and \ref{fig:6} show that the magnetic field strength, angular velocity, and non-axisymmetric perturbation play an important role in fragmentation, and they are closely related.
Thus, it is rather hard  to derive the fragmentation condition from small numbers of calculations.

\subsubsection{Typical Model ($\ap=0.4$)}
\begin{figure}
\begin{center}
\includegraphics[width=160mm]{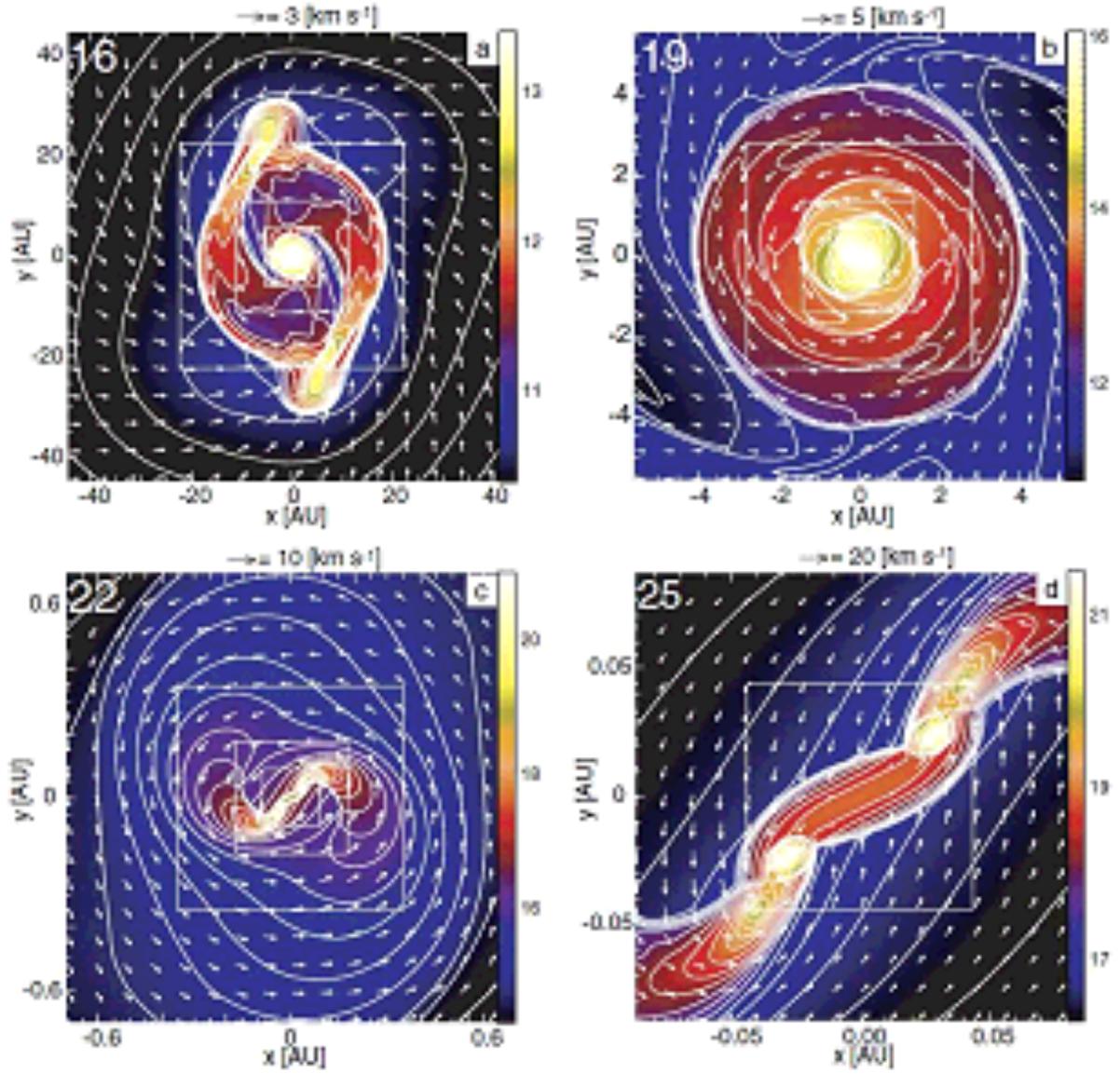}
\caption{
Same as Fig.~\ref{fig:5}.
Final state for model ($\pb$, $\omega$, $\ap$) = (0.01, 0.07, 0.4) is plotted with different grid scales.
}
\label{fig:7}
\end{center}
\end{figure}

Figure~\ref{fig:7} shows the cloud structure on the $z=0$ plane for the model of ($\pb$, $\omega$, $\ap$) = (0.01, 0.07, 0.4).
In this figure, the structure around the center of the cloud is plotted with different scales at the end of the calculation.
The spatial scale of each successive panel differs by a factor of eight and thus the scale between Figures~\ref{fig:7}{\it a} and {\it d} differs by a factor of 512.
In this model, the first core is formed at $\nc \simeq 3\times 10^{12}\cm$.
The first core repeats fragmentation and merger in the adiabatic phase.
As shown in Figure~\ref{fig:7}{\it a}, in a large scale, the high-density region is composed of a central compact core and a surrounding ring-like structure.
At the end of the calculation, the central compact core has the maximum density of $\nc=1.1\times10^{22}\cm$, and a mass of $M=0.065\msun$.
The second collapse occurs in the central core at 459\,yr after the first core-formation epoch.
Then, the central region fragments again at $\nc=8\times 10^{21}\cm$, and two protostars are formed, as shown in Figures~\ref{fig:7}{\it c} and {\it d}.
Two fragments (or protostars) in Figure~\ref{fig:7}{\it d} have a mass of $M=4.4\times 10^{-3}\msun$, and a separation of $R_{\rm sep}=0.073$\,AU at the end of the calculation.

On the other hand, in the surrounding ring-like structure, there are two clumps at ($x$, $y$) $\simeq$ ($\pm$5\,AU, $\mp$25\,AU) as shown in Figure~\ref{fig:7}{\it a}.
These clumps have masses of $M\simeq0.036\msun$.
The clumps have the maximum density of $\nc=8\times 10^{12}\cm$ and they continue to collapse until the calculation ends.
Thus, these clumps are expected to be more compact as time goes on, and then, a single or several protostars are formed in each clump.

As shown in Figure~\ref{fig:7}, four clumps appear in this model.
Clumps formed in the adiabatic phase have a wide separation of $R_{\rm sep}\simeq 26$\,AU, and they seem to be a triple system in a large scale (Fig.~\ref{fig:7}{\it a}).
In the central compact core, there are two fragments, each of which contains a close binary system.
This system is thought to evolve into a hierarchical star system, which is composed of two protostars with a large separation and two binary systems with a narrow separation, if fragments do not merge with each other and separations remain almost unchanged.

\section{DISCUSSION}
\subsection{Initial Cloud Stability and Its Evolution}
In the gravitationally contracting core, the thermal pressure, rotation, and magnetic field work against the self-gravity.
Generally, in the initial clouds, the former  three forces are characterized  by the ratios to the gravitational energy, $\alpha_0$ (thermal energy), $\beta_0$ (rotational energy), and $\gamma_0$ (magnetic energy), and the clouds are specified by the combination of these three parameters.
In the previous section, we investigated the evolutions of clouds with the same $\alpha_0$ ($\alpha_0=0.5$) but different $\beta_0$ and $\gamma_0$.
The magnetic field and rotation  are important for cloud fragmentation, because these forces cause the anisotropic patterns which promote fragmentation.
However, the initial ratio of the thermal to the gravitational energy, which is related to the stability of the initial cloud, may affect cloud fragmentation.
In this section, we compare the evolutions of clouds with the same $\beta_0$ and $\gamma_0$ but different $\alpha_0$.

\begin{figure}
\begin{center}
\includegraphics[width=160mm]{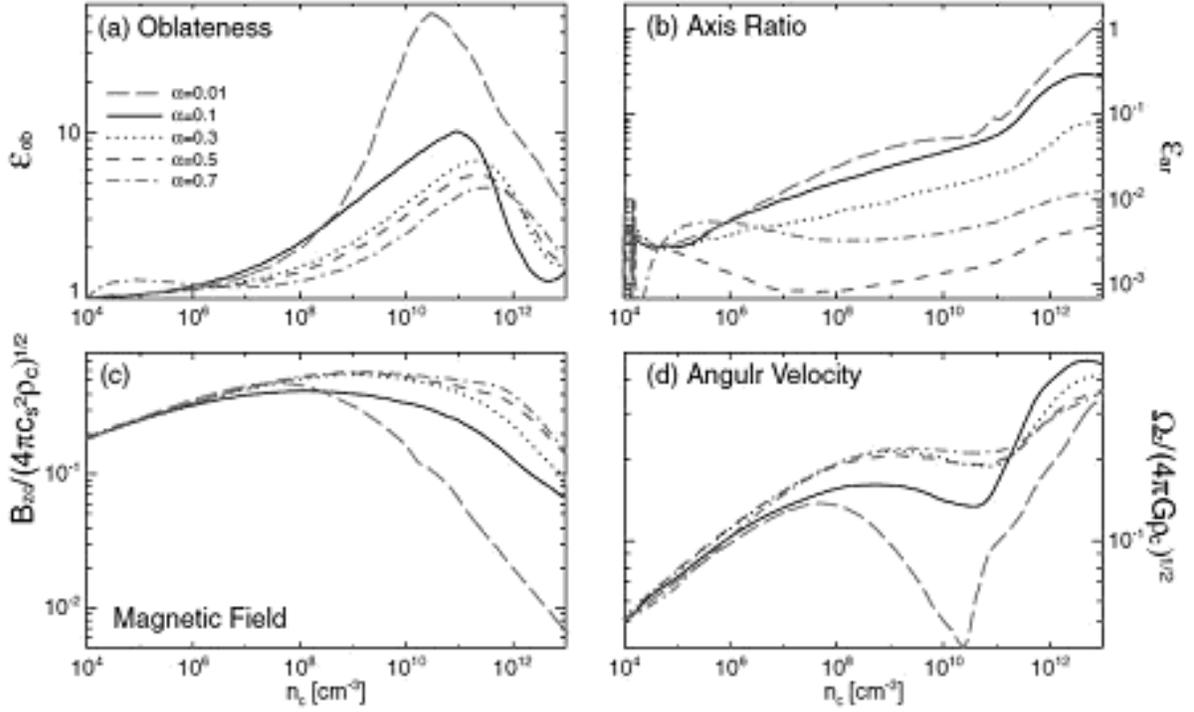}
\caption{
Evolution of models with different $\alpha_0$.
Panel (a): Oblateness $\ob$, (b): axis ratio $\ar$, (c) Magnetic field normalized by the square root of the central density $B_{\rm zc}/(4\pi c_{\rm s,0}^2 \rhoc)^{1/2}$, and (d) angular velocity  normalized by the square root of the central density $\Omega_{\rm c}/(4\pi G \rhoc)^{1/2}$, all of which are plotted against the central density $\nc$.
}
\label{fig:8}
\end{center}
\end{figure}

We investigate the evolutions of clouds with five different $\alpha_0$ ($\alpha_0=$0.01, 0.1, 0.3, 0.5, and 0.7).
The cloud with $\alpha_0=0.01$, which has 1\% of the thermal energy to the gravitational energy at the initial state, is highly gravitationally unstable, while that with $\alpha_0=0.7$ is initially nearly in equilibrium.
We fixed the three other parameters as ($\pb$, $\omega$, $\ap$) = (0.01, 0.05, 0.01).
Figure~\ref{fig:8} shows the evolution of the oblateness $\ob$ (Fig.~\ref{fig:8}{\it a}), axis ratio $\ar$ (Fig.~\ref{fig:8}{\it b}), normalized magnetic field strength $B_{\rm zc}/(4 \pi c_{\rm s,0}^2 \rhoc)^{1/2}$ (Fig.~\ref{fig:8}{\it c}), and normalized angular velocity $\Omega_{\rm zc}/(4\pi G \rhoc)^{1/2}$ (Fig.~\ref{fig:8}{\it d}).

Firstly, we show the evolution of clouds with large $\alpha_0$ ($\alpha_0=$ 0.3, 0.5 and 0.7).
These models are initially more stable.
In Figure~\ref{fig:8}, these models show similar evolutions for oblateness, normalized magnetic field strength, and normalized angular velocity.
After the central densities exceed $\nc \gtrsim 10^{12}\cm$, the central regions have almost the same degrees of oblateness, magnetic fields, and angular velocities as shown in Figures~\ref{fig:8}{\it a}, {\it c}, and {\it d}.
Shock fronts appear  and the first cores surrounded by the shock waves are formed at $\nc \simeq 10^{13}\cm$ in these models.
The first cores have almost the same properties except for the axis ratios.
Comparing all five models, the formed first cores are shown to have larger axis ratios in the models with smaller $\alpha_0$.
\citet{hanawa00} and \citet{lai00} showed that the growth rate of the non-axisymmetric perturbation is larger in a more unstable cloud core from their linear analysis \citep[see also Fig.~14 of][]{omukai05}.
Since the cloud evolution converges to a self-similar solution as the cloud collapses even when the cloud is rotating and magnetized \citep{machida05a,machida06a}, the growth rate of the non-axisymmetric perturbation is also converged in the collapsing cloud.
Thus, the difference in the axis ratio which reflects the amplitudes of the non-axisymmetric perturbation comes from the growth in the early phase of the contraction.
For this reason, the initially more unstable cloud (i.e., the cloud with small $\alpha_0$) has a large amplitude of non-axisymmetric perturbation.

\begin{figure}
\begin{center}
\includegraphics[width=160mm]{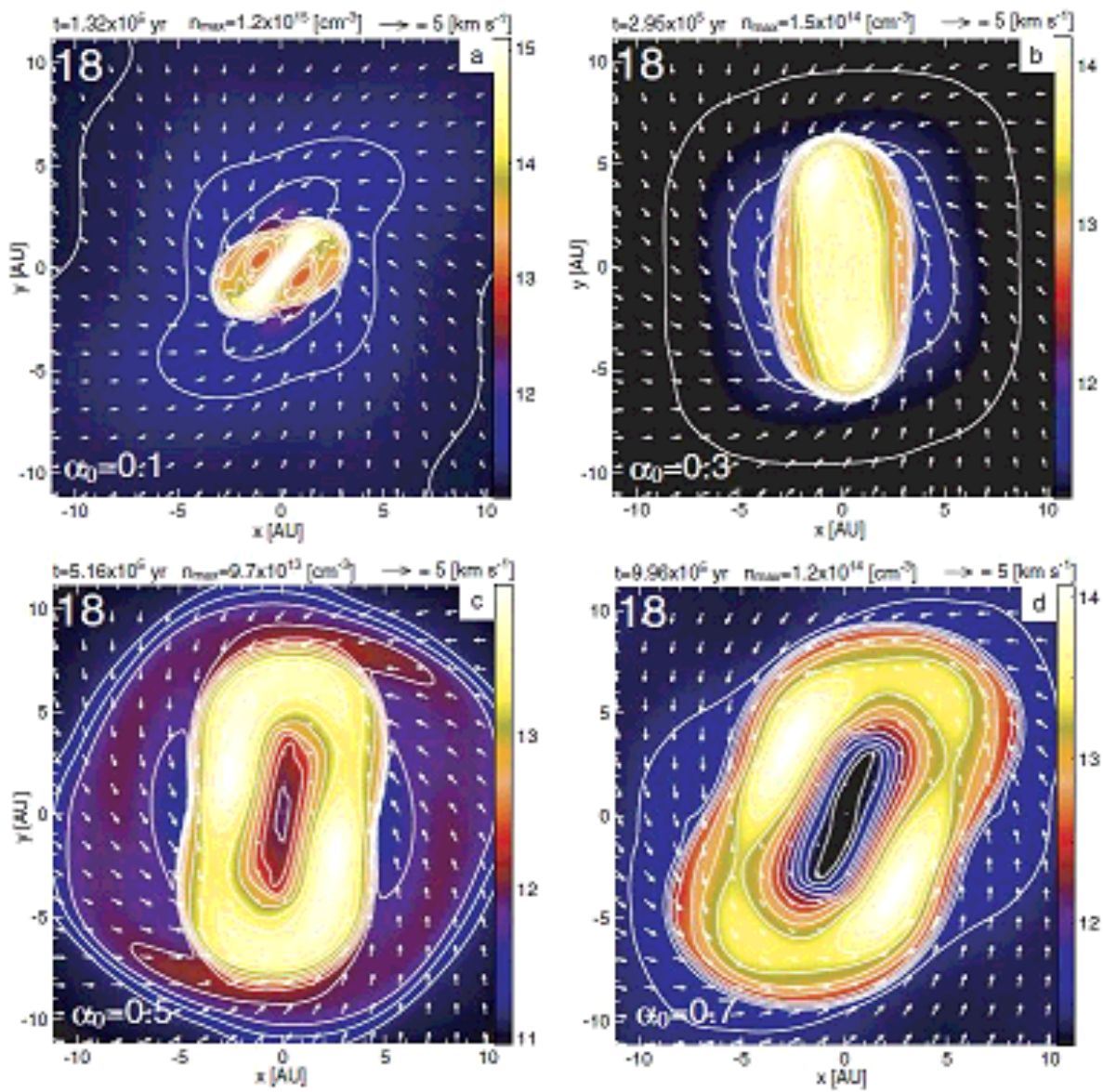}
\caption{
Cloud structures around the center of the cloud for models with different $\alpha_0$ ($\alpha_0=0.1, 0.3, 0.5$, and 0.7) at the fragmentation epoch.
Density (color-scale) and velocity distribution (arrows) on the cross-section in the $z=0$ plane are plotted.
The elapsed time $t$, maximum number density $n_{\rm max}$, and arrow scale are listed in each panel.
The level of the subgrid is shown in the upper left corner of each panel.
}
\label{fig:9}
\end{center}
\end{figure}

Figure~\ref{fig:9} shows the cloud structures near the center of the cloud just after fragmentation.
Fragmentation occurs via a ring-like structure in the models  $\alpha_0=0.5$ and 0.7, while fragmentation occurs without the ring formation in the model with $\alpha_0=0.3$.
This difference in the fragmentation pattern comes from the difference in the axis ratio of each model:  the model with $\alpha_0=0.3$ has a larger amplitude of non-axisymmetric perturbation than those of models with $\alpha_0=0.5$ and 0.7 at the first core formation epoch.
However, the scale and fragmentation epoch are almost the same for all models with $\alpha_0 \gtrsim 0.3$,  as shown in Figures~\ref{fig:9}{\it b}-{\it d}.
Thus, even though the fragmentation patterns are different, the fragmentation scale and its epoch are hardly changed for models $\alpha_0 \gtrsim 0.3$.

Next is shown the evolution of the highly unstable cores with $\alpha_0=0.1$ and 0.01.
Figure~\ref{fig:8} indicates that the evolutions of oblateness, angular velocity and magnetic field of the models with $\alpha_0=0.1$ and 0.01 differ appreciably from those of models with $\alpha_0 \ge 0.3$.
In addition, the first core in the model of $\alpha_0=0.1$ is considerably smaller than those in models with larger $\alpha_0$ (Fig.~\ref{fig:9}).
In the model with $\alpha_0=0.01$, an extremely thin disk with $\ob \simeq 50$ is formed in the isothermal phase (Fig.~\ref{fig:8}{\it a}).
\citet{tsuribe99} expected these thin disks to fragment in the isothermal phase when the cloud collapses isothermally for a long time.
In our calculation, however, fragmentation does not occur in the isothermal phase, because the central region  begins to behave adiabatically just after the oblateness reaches its peak.
Then, an elongated bar is formed in the adiabatic phase, and the bar fragments at the central density $\nc \simeq 10^{15}\cm$.
The fragmentation scale of the model of $\alpha_0=0.01$ is $\sim2$\,AU, which is considerably smaller than in models with $\alpha_0>0.3$.
Thus, fragmentation size becomes smaller with decreasing $\alpha_0$ for $\alpha_0 < 0.3$.

In summary, for models with $\alpha_0>0.3$, the clouds show similar evolutions, and the first cores have almost the same properties.
On the other hand, clouds with $\alpha_0 < 0.3$ evolve in a completely different way, and the first cores have different properties. 
However, observations have shown that molecular cloud cores are observed in nearly thermal equilibrium against gravity  \citep[e.g.,][]{tachihara02}, which indicates $\alpha_0 \sim 1$.
As a result, the evolution of molecular cloud cores is specified by only $\beta_0$ and $\gamma_0$ since clouds with $\alpha_0 > 0.3$ have the same evolutional properties when the clouds have the same $\beta_0$ and $\gamma_0$.

\subsection{Scales and Epochs of Fragmentation}
\begin{figure}
\begin{center}
\includegraphics[width=160mm]{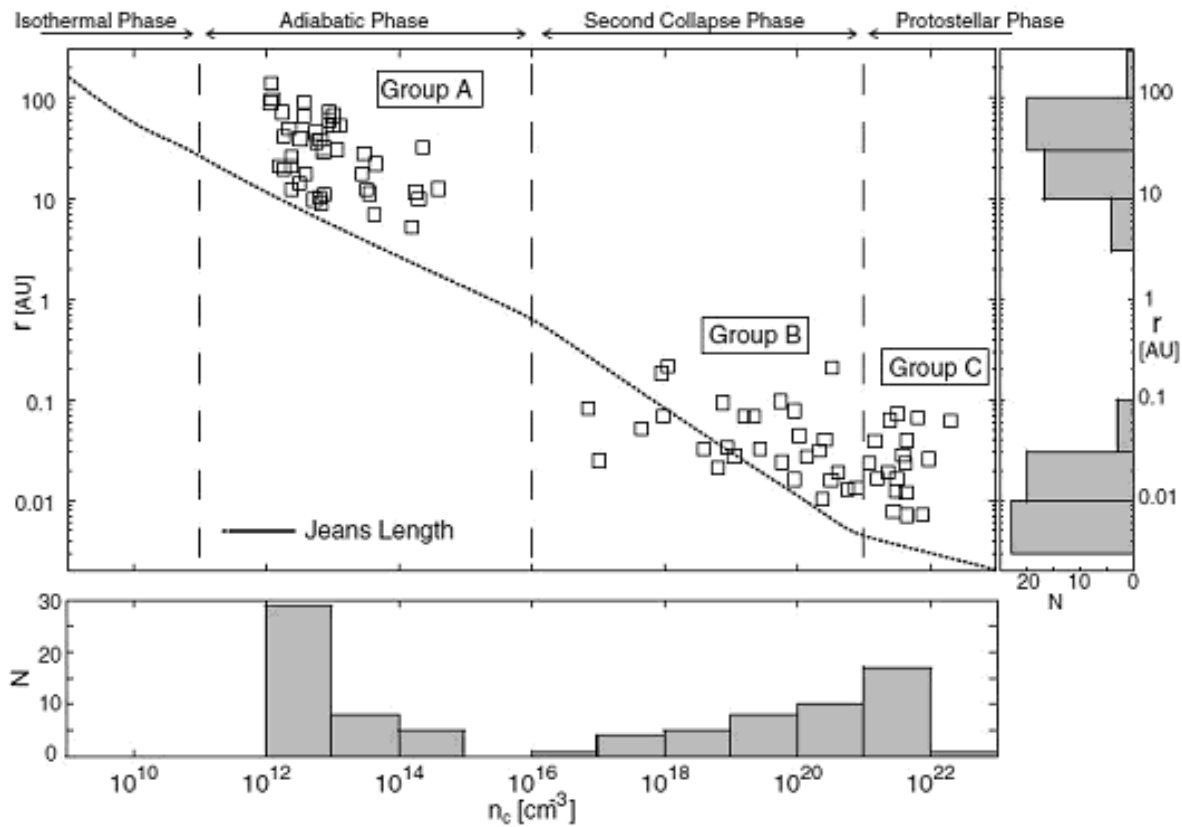}
\caption{
Central panel: Separations between furthermost fragments as a function of the density at the end of the calculation.
The Jeans length evaluated from the spherical symmetric calculations is also illustrated by the dotted line.
The results of all models where fragmentation is observed are plotted.
Models are classified into three groups according to the fragmentation epoch: fragmentation in the adiabatic phase (A), second collapse phase (B), and protostellar phase (C).
Right panel: Histogram of fragmentation models as a function of the separation $r$.
Bottom panel: Histogram of fragmentation models as a function of the fragmentation epoch measured by the central density $n_{\rm c}$. 
}
\label{fig:10}
\end{center}
\end{figure}

We calculated 147 models with different sets of initial cloud rotation, magnetic field, thermal energy, and amplitude of the non-axisymmetric perturbation.
In 102 out of 147 models, we observed fragmentation.
To investigate the scales and epochs of fragmentation statistically, for all models that show fragmentation in Figures~\ref{fig:1}, \ref{fig:4}, and \ref{fig:6}, the number densities at the fragmentation epoch, and the farthermost separations between fragments at the end of the calculation are plotted in Figure~\ref{fig:10} upper left panel (main panel).
Right and lower panels in Figure~\ref{fig:10} show the histogram of fragmentation scales and epochs measured in the central density, respectively.
Although we calculated the cloud evolution from $\nc = 10^4\cm$, we observe no fragmentation in the isothermal phase ($\nc < 10^{11}\cm$).
Thus, only the range of $\nc > 10^{9}\cm$ is plotted.
For convenience, we divide fragmentation models into three groups (Groups A, B, and C) by the evolution stage when fragmentation occurs.
Models in Groups A, B, and C, respectively, show fragmentation in the adiabatic ($10^{11}\cm < \nc < 10^{16}\cm$; A), second collapse ($10^{16}\cm < \nc < 10^{21}\cm$; B), and protostellar phase ($\nc > 10^{21}\cm$; C). 
The dotted line in the main panel shows the Jeans length, which is derived from the relation of the temperature and density assumed in our calculation.
Figure~\ref{fig:10} shows that points are distributed near the dotted line, which indicates that fragments have a separation  nearly equal to the Jeans scale when they are born. 
Note that models are distributed slightly above the Jeans length in Figure~\ref{fig:10}, because a larger critical wavelength is expected for rotating magnetized gas, while the Jeans length plotted here is expected to be a spherical symmetry.

Figure~\ref{fig:10} shows that there are two distinct fragmentation epochs, namely  $\nc \simeq 10^{12}\cm$ and $\nc \simeq 10^{21}\cm$.
The lower panel shows models belonging to groups B and C to be smoothly distributed, but there is a clear gap between groups A and B for $10^{15}\cm \lesssim \nc \lesssim 10^{16}\cm$.
Models in group A are distributed only in the range of $10^{12}\cm<\nc < 10^{15}\cm$, although the adiabatic phase lasts for $10^{11}\cm < \nc < 10^{16}\cm$.
In group A, the fragmentation epochs are bunched into the first core formation epoch.
The first core is formed at lower density when the host cloud has a strong magnetic field or rapid rotation.
The central density at the first core formation epoch is distributed for $3\times 10^{11}\cm \lesssim \nc \lesssim 10^{15}\cm$.
In group A, models that fragment at relatively low density have strong magnetic fields or rapid rotations, while models that fragment at relatively high density have weak magnetic fields and slow rotations.
In addition, no fragmentation is expected for $10^{15}\cm < \nc < 10^{16}\cm$.
This means that the core fragments just after the first core formation, and fragmentation never occurs if the core fails at this epoch.

In group A, fragments have mutual separations of $3-300$\,AU.
Since we stopped calculation in group A after fragmentation occurs,  we do not follow the cloud evolution until the protostar is formed ($\nc \simeq 10^{21}\cm$).
Thus, fragmentation may occur again for these models in the second collapse or protostellar phase.
In that case, it is expected that hierarchical stellar systems are formed.

Some models that do not show fragmentation in the adiabatic phase fragment in the second collapse phase (group B).
The lower panel in Figure~\ref{fig:10} shows that the number of fragmentation models smoothly increases with the cloud density for $\nc \gtrsim 10^{16}\cm$.
As shown in Figures~\ref{fig:3} and \ref{fig:8}, the central region has a considerably weak magnetic field in the second collapse phase owing to the Ohmic dissipation \citep[for detail, see][]{machida07a}.
Thus, the magnetic braking is not so efficient that the central region can spin up as the cloud collapses.
Due to this accelerated rotation, fragmentation is induced in the second collapse phase.
Fragments in group B have a typical separation of $0.01-0.3$\, AU.

When the central density exceeds $\nc \gtrsim 10^{21}\cm$,  the second core (or protostar) is formed.
Even after the second core formation, fragmentation frequently occurs.
In group C, fragmentation occurs just after the second core formation, similar to the fragmentation in the adiabatic phase.
Fragments in group C have a typical separation of $0.005-0.07$\,AU.

The fragmentation process both in the adiabatic and protostellar phases is thought to differ from that in the second collapse phase.
Both in the adiabatic and protostellar phases, fragmentation occurs after the respective cores are formed.
Fragmentation easily occurs in a quasi-static core, and the core has enough time to amplify the perturbation that induces fragmentation.
On the other hand, fragmentation occurs owing to cloud rotation in the second collapse phase.
Fragmentation does not occur in the isothermal phase when the cloud collapses in a self-similar fashion, because the self-similar solution with $\gamma = 1$ exists even in a rotating collapsing cloud \citep{matsu97,matsu99}. 
On the other hand, the clouds with $\gamma>1$ form rotating disks and stop the contraction, because no self-similar solution exists \citep{saigo00}.
These clouds can fragment as shown in \citet{saigo04}.
In addition, since the cloud collapses with $\gamma=1.1$, the non-axisymmetric perturbation manages to evolve in this phase as shown in \citet{hanawa00} and \citet{lai00}.
Thus, the fragmentation process in groups A and C is different from group B, which causes a gap between group A and group B.
However we cannot distinguish group B from group C, because the number of fragmentation models smoothly increases after $\nc \gtrsim 10^{16}\cm$.
As a result, the densities when fragments appear are divided into two groups:  $10^{12} \cm\lesssim  \nc \lesssim  10^{15}\cm$ and $\nc\gtrsim 10^{16}\cm$.
Histogram of separation shows that the separations between fragments are clearly divided into two groups (Fig.~\ref{fig:10} right panel): models anticipating wide ($3-300$\,AU) separations and those anticipating narrow ($0.005-0.3$\,AU) separations.
However, this feature of two distinct groups of separation may be smoothed out as long as the protostar evolves.
In addition, we stopped the calculation after the first fragmentation occurs or the protostar is formed.
Thus, further fragmentation (or protostellar formation) may occur around a protostellar disk and the first core. 
Thus, although further lengthy calculations are needed to know the real distribution of binary separations, we may observe two distinct groups of binary in young clusters composed of extremely young stars.

\subsection{Does Magnetic Field Suppress or Promote Fragmentation?}

\begin{figure}
\begin{center}

\includegraphics[width=160mm]{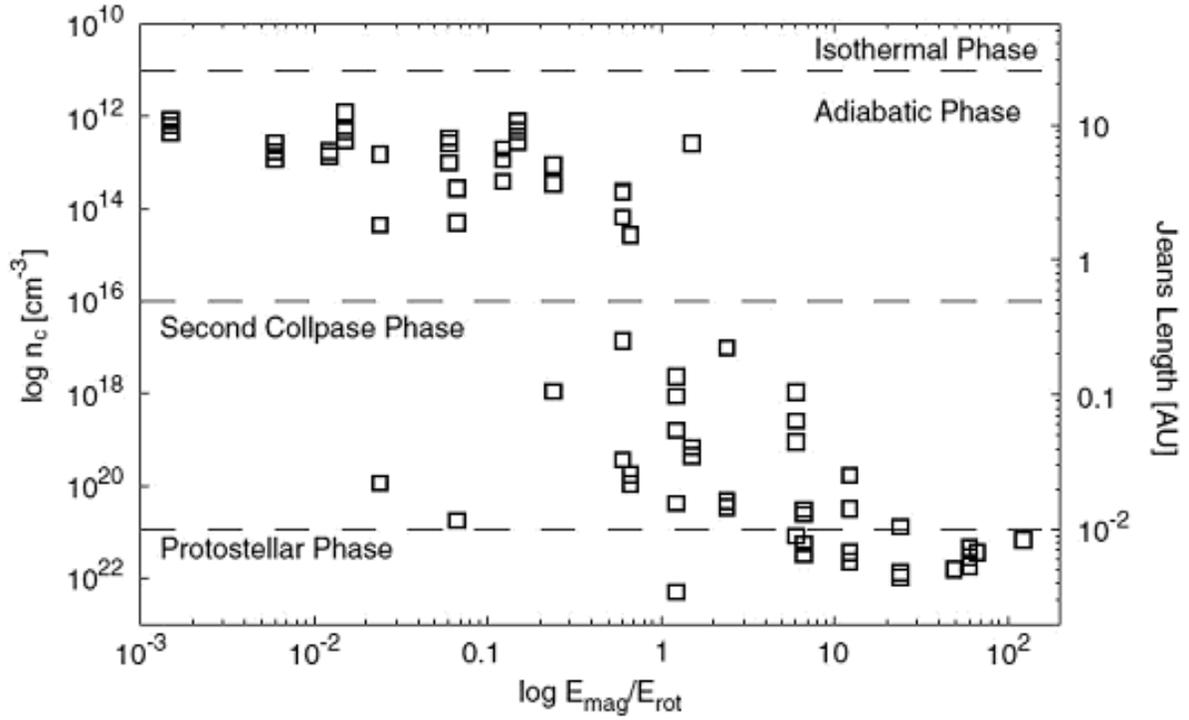}
\caption{
Number density at the fragmentation epoch as a function of the ratio of the magnetic to the rotation energies $E_{\rm mag}/E_{\rm rot}$ of the initial cloud.
Right axis indicates the Jeans length.
The dashed lines mean the borders between different evolutional phases.
}
\label{fig:11}
\end{center}
\end{figure}

In Figure~\ref{fig:11}, the fragmentation epochs are plotted against the ratio of the initial magnetic energy to the rotational energy ($E_{\rm mag}/E_{\rm rot}$).
The right axis in this figure means the Jeans length that is calculated from the number density (right axis), in which the temperature is related to the number density by the spherical symmetric calculation \citep{masunaga00}.
Only fragmentation models in the magnetized clouds are plotted in this figure.
The figure shows that the fragmentation epochs shift to high density as $E_{\rm mag}/E_{\rm rot}$ increases.
This indicates that cloud rotation promotes fragmentation, while the magnetic field suppresses fragmentation.
This is valid through all the phases of cloud evolution from the isothermal to the protostellar phases.
When the magnetic energy is lower than the rotational energy $E_{\rm mag}<E_{\rm rot}$ in the initial cloud, many models result in fragmentation in the adiabatic phase.
On the other hand, when the magnetic energy is predominant over the rotational energy $E_{\rm mag}>E_{\rm rot}$, fragmentation occurs mainly in the second collapse and protostellar phases, and some models do not show fragmentation.
In this figure, there are some exceptions.
Two models show fragmentation in the second collapse phase even if $E_{\rm mag}/E_{\rm rot}<0.1$.
In these models, fragmentation does not occur in the adiabatic phase, because the non-axisymmetric perturbation is moderately grown and an induced spiral pattern effectively removes the angular momentum from the center of the cloud.
Thus, fragmentation is postponed.

Observations have shown that molecular cloud cores have high magnetic energies, but low rotational energies \citep{crutcher99,caselli02}, which indicates $E_{\rm mag}/E_{\rm rot} \gg 1$
Figure~\ref{fig:11} indicates that these clouds barely fragment.
This conflicts with observations that the majority of stars are born as binary \citep[e.g.,][]{duq91}.
In this study, we calculated the evolution of a magnetized cloud in which the initial magnetic field lines are aligned to the rotation axis.
When magnetic field lines are not aligned to the rotation axis, cloud evolution may be changed.
However, \citet{machida06a} showed that when the magnetic field lines are not aligned to the rotation axis at the initial state, the magnetic braking is more effective, and the angular momentum is effectively transferred outwardly. 
This implies that fragmentation is more suppressed.
On the other hand, \citet{price07} showed, in their ideal MHD calculations, that when initial cloud has a considerably distorted structure, fragmentation can occur even in a strongly magnetized cloud.

To understand observational binary frequency,  we may need to calculate the cloud evolution in a non-ideal regime, parameterizing the cloud shape and the angle between the rotation axis and the magnetic field lines, besides the non-axisymmetric perturbation, and the magnetic, rotational, and thermal energies.

\subsection{Fragmentation Condition and Wide/Close Binary Formation}
\begin{figure}
\begin{center}
\vspace{-2cm}
\includegraphics[width=140mm]{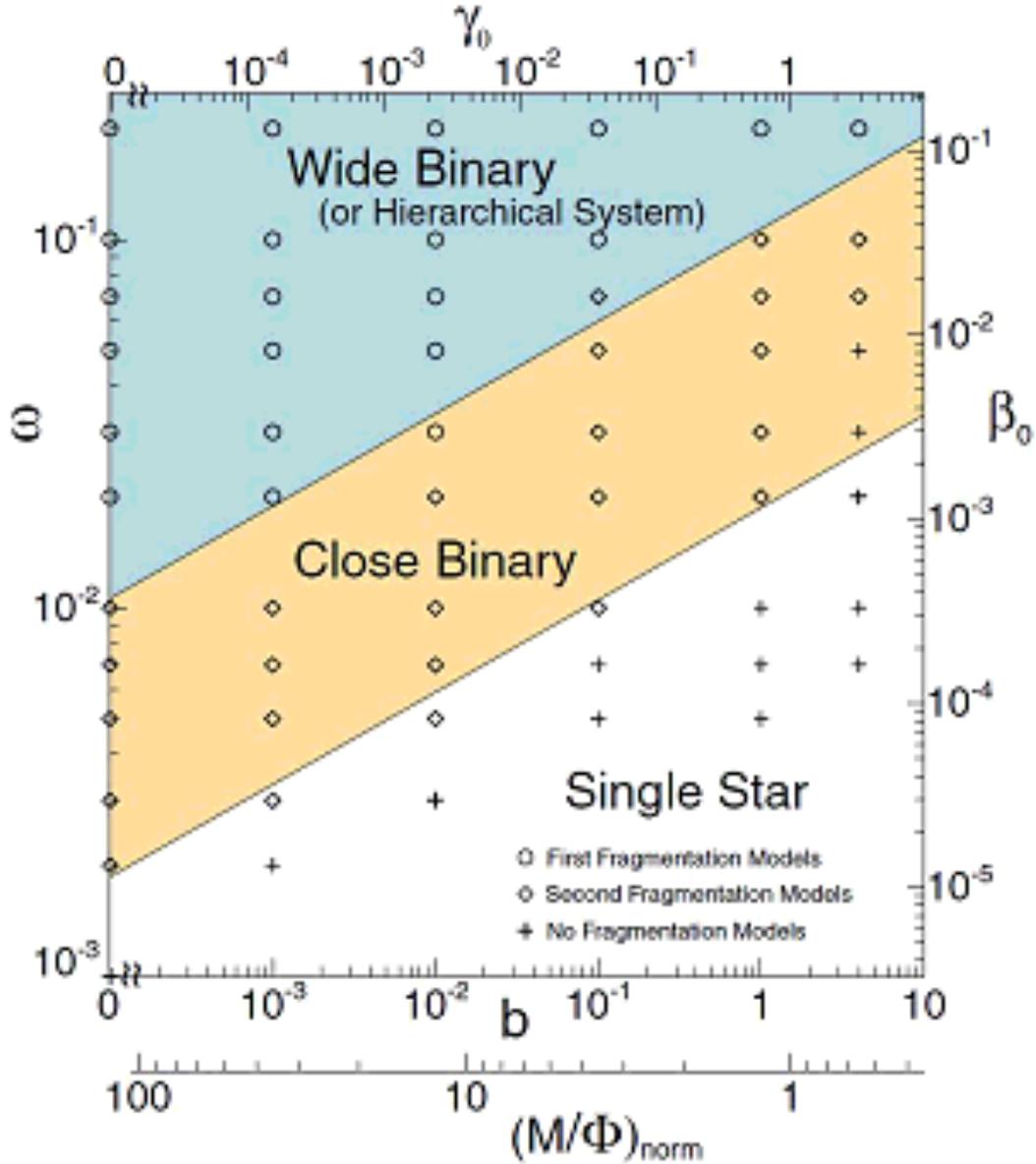}
\caption{
Fragmentation condition against parameters $\pb$ and $\omega$.
Symbols mean that circle $\circ$: fragmentation occurs in the adiabatic phase, diamond $\diamond$: fragmentation occurs in the second collapse phase, cross $+$: no fragmentation occurs through all phases of the cloud evolution.
Models in the wide and close binary regions have separations in the range of $3-300$\,AU (wide binary region ), and $0.007-0.3$\,AU (close binary region).
Models in the single-star region do not show fragmentation in any evolution phase.
Upper and right axis indicate the ratio of the magnetic $\gamma$ (upper) and rotational $\beta$ (right) energies to the gravitational energy.
Bottom axis means the mass-to-magnetic flux ratio M/$\Phi$ normalized by the critical value (see, Eq~\ref{eq:crit}).
}
\label{fig:12}
\end{center}
\end{figure}

We investigated the cloud evolutions controlled by four parameters.
Each parameter corresponds to the magnetic $\pb$, rotational $\omega$, thermal energies $\alpha_0$, and the amplitude of the non-axisymmetric perturbation $\ap$ of the initial cloud.
As shown in the previous sections, fragmentation significantly depends on the magnetic field strength and rotation rate of the initial cloud, while it slightly depends on the thermal energy and the amplitude of the non-axisymmetric perturbation.
Thus, in Figure~\ref{fig:12}, we have plotted the final state of each cloud against the magnetic field strength and rotation rate of the initial cloud.
In this figure, $x$- and $y$-axes are parameters $\pb$ and $\omega$, respectively.
Upper and right axes mean, respectively, the ratio of the magnetic ($\gamma_0$) and rotational ($\beta_0$) energies to the gravitational energy of the initial cloud.
Bottom axis means the mass to magnetic flux ratio $(M/\Phi)_{\rm norm}$ normalized by the critical value (see, Eqs~10--12).
In this figure, some models are located in the region of $\gamma_0 > 1$ or $(M/\Phi)_{\rm norm}<1$, which indicates that the cloud is  magnetically subcritical.
In these models, the initial cloud is magnetically subcritical as a whole, while the central region is magnetically supercritical since we adopted the Bonnor-Ebert density profile \citep[for detail, see][]{machida07a}. 
Thus, any cloud can collapse in a sufficiently shorter timescale than the ambipolar diffusion timescale.

In Figure~\ref{fig:12}, three types of symbols (circle $\circ$, diamond $\diamond$, and cross $+$) are plotted against the initial magnetic fields and angular velocities.
Each symbol includes three models  with different $\ap$ ($\ap=0.01$, 0.2, and 0.4). 
Symbols indicate that (i) Circle $\circ$: At least one model out of three shows fragmentation in the adiabatic phase ($10^{11}\cm \lesssim \nc \lesssim 10^{16}\cm$), (ii) Diamond $\diamond$: in models that do not show  fragmentation in the adiabatic phase ($\nc \lesssim 10^{16}\cm$),  at least one model out of three shows fragmentation in the second collapse or protostellar phase ($\nc \gtrsim 10^{16}\cm$), and (iii) Cross $+$: models show no fragmentation through any phases of cloud evolution  ($10^4\cm<\nc < 10^{22}\cm$).  
As shown in Figure~\ref{fig:12}, we can divide the parameter space into three regions: wide fragmentation, close fragmentation, and non-fragmentation, which correspond to the formation of wide binary, close binary, and single star.
The separations between fragments are clearly divided into two classes (wide and narrow separation) according to the fragmentation epoch, as shown in Figure~\ref{fig:10}.

Figure~\ref{fig:12} indicates that models with strong magnetic fields and rapid rotations are distributed in the wide binary region.
Models in the wide binary region {\em can} fragment in the adiabatic phase, and fragments have separations of $3-300$\,AU.
However, not all models in this region fragment in the adiabatic phase.
Even among models in this region, some models show fragmentation only in the second collapse or protostellar phase, and some models show no fragmentation through any phases of cloud evolution.
For example, models ($\pb$, $\omega$) = (0.01, 0.05), which are located in the wide binary region in Figure~\ref{fig:12}, show fragmentation in the adiabatic phase for $\ap=0.01$ (Fig.~\ref{fig:1}) and 0.4 (Fig.~\ref{fig:6}), while fragmentation occurs only in the second collapse phase for the model with $\ap=0.2$ (Fig.~\ref{fig:4}).
In models ($\pb$, $\omega$) = (0.001, 0.07), fragmentation occurs in the adiabatic phase for the models with $\ap=0.01$ (Fig.~\ref{fig:1}) and $0.2$ (Fig.~\ref{fig:4}), while the protostar is directly formed without fragmentation for the model with $\ap=0.4$ (Fig.~\ref{fig:6}).
Thus, a small difference in the amplitude of the non-axisymmetric perturbation at the initial state has a possibility to induce a large difference in the fragmentation epoch. 
In some models located in the wide binary region, we stopped calculation before the protostar was formed when the fragmentation occurred in the adiabatic phase, because the fragments are likely to escape from the finest grid.
Thus,  fragmentation can occur again in each fragment in the second collapse or  protostellar phase, which seems to lead to a hierarchical stellar system.

Models distributed in the close binary region have a weaker magnetic field and slower rotation than models in the wide binary region.
However, the tendency of rotation to promote fragmentation, and the magnetic field to suppress fragmentation is the same.
Models in the close binary region {\em can} fragment in the second collapse or protostellar phase, and fragments have separations of 0.005--0.3\,AU as shown in Figure~\ref{fig:10}.
However, similarly to wide binary models, even in models in the close binary region, some models do not fragment in either the second collapse or protostellar phases.
For example, in models ($\pb$, $\omega$) = (1, 0.05), fragmentation occurs in the protostellar phase for the model with $\ap=0.01$ (Fig.~\ref{fig:1}), while protostars are directly formed without fragmentation for the models with $\ap=0.1$ (Fig.~\ref{fig:4}), and 0.4 (Fig.~\ref{fig:6}).

Models in the single-star region have stronger magnetic fields and slower rotations than those in the wide and close binary regions.
In all models in this region, the protostar is directly formed without fragmentation.
As a result, a single compact sphere is formed at the center of the cloud, as shown in models ($\pb$, $\omega$) = (4, 0.007) located in the lowest right corner of Figures~\ref{fig:1}, \ref{fig:4}, and \ref{fig:6}.
These compact spheres are expected to evolve into single stars.

Models located in the wide/close binary regions have the possibility of forming the wide/close binary.
As for fragmentation, it is difficult to forecast the cloud evolution from the initial conditions, and a small difference in the initial state sometimes results in different final outcomes.
For example, even models located in the wide binary region have three possibilities: formation of a wide binary, a close binary, or a single star.
That is, the parameters in wide/close binary regions are the necessary conditions, not sufficient conditions for wide/close binary formation.

\subsection{Comparison with Observation}
As shown in Figures~\ref{fig:11} and \ref{fig:12}, our results indicate that cloud rotation promotes fragmentation, but the magnetic field suppresses fragmentation.
In this section, to quantify the magnetic field and cloud rotation in the molecular cloud core, we compare our results with observations.

Both magnetic field and rotation rate are observed in molecular cloud TMC-1C.
TMC-1C has $\beta = 1.2\times 10^{-3}$ \citep{goodman93}, and $\gamma < 0.08 $ \citep{crutcher99}.
As for magnetic field strength, only maximum value was estimated for observational limitation.
When a cloud has $\beta = 1.2\times 10^{-3}$, Figure~\ref{fig:12} indicates that (i) fragmentation occurs in the adiabatic phase and the wide binary can form when $\gamma < 10^{-3}$, (ii) fragmentation occurs either in the second collapse or protostellar phases, and the close binary can form when $10^{-3} < \gamma < 0.45$, and (iii) only a single star forms without fragmentation when $\gamma > 0.45$.
Since observation showed $\gamma < 0.08$, it is possible to form a binary system in TMC-1C.

In the observations of molecular cloud core, \citet{crutcher99} showed that the magnetic energy is comparable to gravitational energy ($\gamma\sim0.5$), while \citet{goodman93} and \citet{caselli02} showed that rotational is much smaller than gravitational energy ($\beta\sim0.02$).
Thus, these observations indicate that it is difficult to form a (wide) binary system as shown in \S5.3.
Except for TMC-1C, however, there are few clouds in which both magnetic field and rotation rate are observed.
In addition, the observed magnetic  and rotation energies might not reflect the distribution of these energies in the majority of cloud cores, because we can only observe a limited range of the magnetic field strength and rotation speed: we cannot observe weak magnetic field strengths and slow rotation speeds.
Furthermore, we have to measure the magnetic field strength and rotation rate at the same radius or the same number density in the molecular cloud core.
To determine magnetic field strength and rotation rate precisely, we need a high-resolution observational facility such as the future ALMA.

\subsection{Comparison with Previous Work}
Fragmentation conditions in the adiabatic phase (i.e., the wide binary region) were investigated by \citet{machida05b}, and \citet{price07} using the ideal MHD approximation, which is not valid in the high density gas region with $10^{12} \lesssim \nc \lesssim 10^{16}\cm$.
In this paper, we studied cloud evolution in the non-ideal MHD regime.
However, the wide binary region in Figure~\ref{fig:12} corresponds well to results of \citet{machida05b}.
As shown in Figure~\ref{fig:10}, many models located in the wide binary region fragment in the range of $10^{12}\cm \lesssim \nc \lesssim 10^{13}\cm$. 
In addition, since the density outside a small central part of the cloud is $\nc \lesssim 10^{12}\cm$, the Ohmic dissipation is not so effective at the fragmentation epoch.
From this, there must be only a little difference between the ideal MHD and non-ideal MHD calculations for fragmentation appearing in the adiabatic phase.

\citet{price07} investigated the evolution of magnetized clouds and fragmentation conditions with their MHD-SPH code.
Their results are qualitatively the same as ours: the magnetic field suppresses fragmentation.
However, there is a small difference.
They adopted a rotation energy of $\beta_0=0.005$ for the initial clouds.
Fragmentation occurs when $(M/\Phi)_{\rm norm} > 5$ in \citet{price07}, while fragmentation occurs when $(M/\Phi)_{\rm norm} > 10$ in our calculations for models with $\beta_0=0.005$, as shown in Figure~\ref{fig:12}.
Thus, in \citet{price07}, fragmentation occurs in stronger magnetic fields than ours.

Fragmentation is promoted by  rotation, but it is suppressed by the magnetic field.
In strongly magnetized clouds, since the cloud rotation is removed from the central region by the magnetic braking and outflow, fragmentation rarely occurs. 
Outflow is closely related to the magnetic braking: both outflow and magnetic braking are caused by the torsional Alfv\'en wave or the magnetic tension force generated by rotation of the  central core.
Thus, the outflow is one of the proofs that the Alfv\'en wave is properly resolved.
When the Alfv\'en waves are resolved, the outflow always appears after the adiabatic core (or first core) is formed.
In the collapsing clouds, no outflow appears in \citet{price07}, while the outflow appears in the grid-based MHD simulations, such as \citet{tomisaka98,tomisaka02}, \citet{matsu04}, \citet{machida04,machida05b}, \citet{fromang06}, and \citet{banerjee06}.
It is expected that the angular momentum transfer caused by the torsional Alfv\'en wave (or the magnetic tension force) may not be correctly resolved in \citet{price07}, thus excess angular momentum around the center of the cloud is left.
Therefore, in their calculation, fragmentation occurs more frequently than ours.

\section{SUMMARY}
We investigate the cloud evolutions for $10^4\cm < \nc \lesssim 10^{22}\cm$ in a large parameter space.
In this study, we systematically calculated  147 models with different magnetic field, rotation, thermal energies, and the amplitude of the non-axisymmetric perturbation of the initial cloud.
These calculations indicate that 
\begin{enumerate}
\item Fragmentation significantly depends on the magnetic field and rotation, but slightly depends on the thermal energy and the amplitude of the non-axisymmetric perturbation of the initial cloud.
\item The magnetic field suppresses fragmentation, and rotation promotes fragmentation through all phases of cloud evolution.
\item The distribution of the separations between fragments are clearly divided into two classes: fragments formed in the adiabatic phase have wide separations  as $3-300$\,AU, while fragments formed in the second collapse and protostellar phases have narrow separations as $0.007-0.3$\,AU.
\end{enumerate}

\acknowledgments
We have greatly benefited from discussion with ~T. Nakano, ~T. Tsuribe, and ~K. Saigo.
We also thank T. Hanawa for contributing to the nested grid code.
Numerical computations were carried out on VPP5000 at the Center for Computational Astrophysics, CfCA, of the National Astronomical Observatory of Japan.
This work is supported by the Grant-in-Aid for the 21st Century COE "Center for Diversity and Universality in Physics" from the Ministry of Education, Culture, Sports, Science and Technology (MEXT) of Japan, and partially supported by 
the Grants-in-Aid from MEXT (15740118, 16077202,18740113, 18740104).

\begin{table}  
\caption{Model parameters of the magnetic fields}
\label{table:alpha}
\begin{center}
\begin{tabular}{c|ccc} \hline
$\pb$ &  $\gamma_0$ &  $B_0$ ($\mu$G)  & $(M/\Phi)_{\rm norm}$\\ \hline
   0     & 0 &  0  & $\infty$ \\
   0.001 & 9.59$\times10^{-4}$ & 0.5407&39.1   \\
   0.01  & 9.59$\times10^{-3}$ & 1.71& 12.4 \\
   0.1   & 9.59$\times10^{-2}$ & 5.41 &3.91  \\
   1     & 9.59$\times10^{-2}$ & 17.1 &1.24  \\
   4     & 3.83 & 34.1 & 0.62  \\
\hline
\end{tabular}
\end{center}
\end{table}
\begin{table}  
\caption{Model parameters of the angular velocities}
\label{table:omega}
\begin{center}
\begin{tabular}{c|cc} \hline
$\omega$ &  $\beta_0$ &  $\Omega_0$ (10$^{-14}$s$^{-1}$) \\ \hline
   0.003 & 2.97$\times 10^{-5}$ & 0.07   \\
   0.005 & 8.24$\times 10^{-5}$ & 0.116  \\
   0.007 & 1.62$\times 10^{-4}$ & 0.163  \\
   0.01  & 3.30$\times 10^{-4}$ & 0.233  \\
   0.03  & 2.97$\times 10^{-3}$ & 0.698  \\
   0.05  & 8.24$\times 10^{-3}$ & 1.16   \\
   0.07  & 1.62$\times 10^{-2}$ & 1.63   \\
   0.1   & 3.30$\times 10^{-2}$ & 2.33   \\
   0.2   & 0.132                & 4.65   \\
\hline
\end{tabular}
\end{center}
\end{table}

\clearpage

\end{document}